\documentclass[sigconf]{acmart}

\usepackage{url}

\renewcommand\footnotetextcopyrightpermission[1]{Published in IEEE Transactions on Knowledge and Data Engineering (Volume: PP, Issue: 99). Available at \url{https://doi.org/10.1109/TKDE.2017.2747552}.} % removes footnote with conference information in first column
\begin{document}

\title{Beautiful and damned. Combined effect of content quality and social ties on user engagement}

\author{Luca~M.~Aiello}
\affiliation{
  \institution{Nokia Bell Labs}
}
\email{luca.aiello@nokia-bell-labs.com}

\author{Rossano~Schifanella}
\affiliation{
  \institution{University of Turin}
}
\email{schifane@di.unito.it}

\author{Miriam~Redi}
\affiliation{
  \institution{Nokia Bell Labs}
}
\email{miriam.redi@nokia-bell-labs.com}

\author{Stacey~Svetlichnaya}
\affiliation{
  \institution{Flickr}
}
\email{stacey@yahoo-inc.com}

\author{Frank~Liu}
\affiliation{
  \institution{Flickr}
}
\email{frank@yahoo-inc.com}

\author{Simon~Osindero}
\affiliation{
  \institution{Flickr}
}
\email{simon@yahoo-inc.com}

\begin{abstract}
User participation in online communities is driven by the intertwinement of the social network structure with the crowd-generated content that flows along its links. These aspects are rarely explored jointly and at scale. By looking at how users generate and access pictures of varying beauty on Flickr, we investigate how the production of quality impacts the dynamics of online social systems. We develop a deep learning computer vision model to score images according to their aesthetic value and we validate its output through crowdsourcing. By applying it to over 15B Flickr photos, we study for the first time how image beauty is distributed over a large-scale social system. Beautiful images are evenly distributed in the network, although only a small core of people get social recognition for them. To study the impact of exposure to quality on user engagement, we set up matching experiments aimed at detecting causality from observational data. Exposure to beauty is double-edged: following people who produce high-quality content increases one's probability of uploading better photos; however, an excessive imbalance between the quality generated by a user and the user's neighbors leads to a decline in engagement. Our analysis has practical implications for improving link recommender systems.
\end{abstract}

\maketitle

%===========================================
%===========================================
\section{Introduction} \label{sec:intro}
%===========================================
%===========================================

The user experience in online communities is mainly determined by the social network structure and by the user-generated content that members share through their social connections. The relationship between social network dynamics and user experience~\cite{halvey07exploring,susarla12social}, as well as the influence of quality of content consumed on user engagement~\cite{bouch00quality,ceaparu04determining,gulliver06defining} have been extensively researched. However, the relationship between network properties and the production of quality content remains largely unexplored. This interplay is key to reach a full understanding of the user experience in online social systems. Learning how people engage with a platform in relation with the content they produce and consume is crucial to prevent churning of existing users, keep them happy, and attract newcomers.

The growing availability of interaction data from social media, along with the development of increasingly accurate computational methods to evaluate quality of textual and visual content~\cite{luo2014rapid,jin16deep,kong16photo,mai16composition}, has recently provided effective means to fill this knowledge gap. We tap into this opportunity and we aim to advance this research direction by providing the first large-scale study on the production and consumption of quality in online social networks. 

We do so through three main contributions. First, we develop a new deep learning model able to capture the beauty of a picture ($\S$\ref{sec:aesthetics}), as confirmed by a large-scale human crowdsourcing evaluation ($\S$\ref{sec:crowdsourcing}). Second, by applying the model to 15B public photos from Flickr ($\S$\ref{sec:dataset}), we are able to draw the quality profile of the photo collections uploaded by several million users and to partition these users into coherent classes based on the combination of their connectivity, popularity, and contributed quality. This provides the largest-scale description to date of the distribution of quality in an online community. We explore for the first time the relationship between quality production and network structure ($\S$\ref{sec:network}). Most importantly, we set up matching experiments aimed at inferring causal relationships from longitudinal data which allows us to learn more about the combined effect of social network connectivity and the process of quality production on user behavior.

Key findings from the analysis include the following:
\begin{itemize}
\item Unlike popularity, quality is evenly distributed across the network. The resulting mismatch between talent and attention received leaves large portions of the most proficient users with little peer recognition. Users who produce high-quality content but receive little social feedback tend to stay active only for short periods.
\item The level of user-generated quality is correlated with individual social connectivity, which causes a majority illusion effect: users are exposed to images whose average beauty is considerably higher than the average beauty of photos in the platform.
\item Users tend to be assortatively connected with others who produce pictures with similar beauty levels to their own. We find that this network property is partly credited to influence (following talented people increases one's content beauty in the near future) and by the instability of social connections with high imbalance of contributed qualities (users tend to become inactive or churn out if the quality of their neighbors' photos is substantially higher or lower than their own).
\end{itemize}

\noindent The outcomes of our study have practical implications in the domain of recommender systems. We sketch a simple proof-of-concept of a social link recommender algorithm that maximizes the beauty flow while limiting the beauty imbalance between friends ($\S$\ref{sec:application}). Simulations show that this simple strategy balances beauty supply and demand, increasing the level of social inclusion in the class of talented yet unpopular users.

%===========================================
%===========================================
\section{Related Work} \label{sec:related}
%===========================================
%===========================================

\vspace{5pt} \noindent \textbf{Computational Aesthetics.} 
With this work, we build on recent literature exploring the possibility of measuring the intrinsic visual quality of images. Previous related work belongs to the research field of \textit{computational aesthetics}, a domain in which computer vision is used to estimate image beauty and quality. Traditional aesthetic prediction methods are based on handcrafted features reflecting the compositional characteristics of an image. Datta et al.~\cite{datta} and Ke et al.~\cite{ke2006design} were pioneers in this field, with their early work on training classifiers to distinguish amateur from professional photos. Researchers have produced increasingly more accurate aesthetic models by using more sophisticated visual features and attributes~\cite{nishiyama2011aesthetic,dhar11high}, looking at the contribution of semantic features~\cite{marchesotti2011assessing,murray2012ava}, and applying topic-specific models~\cite{luo08photo, obrador09role} and aesthetic-specific learning frameworks~\cite{wu2011learning}. Similar hand-crafted features have successfully been employed to predict higher-level visual properties, such as image affective value~\cite{emotions}, image memorability~\cite{isola2011}, video creativity~\cite{redi6}, and video interestingness~\cite{redi12where, jiang2013understanding}. Such hand-engineered features are of crucial importance for computer vision frameworks requiring interpretability. Recently, Convolutional Neural Networks (CNNs) have become a very popular alternative to hand-crafted features in the computer vision domain, due to their impressive performance on image analysis tasks~\cite{ilscrvc}. The few pieces of work that tested CNNs for aesthetic scoring have done so on professional image corpora~\cite{luo2014rapid,mai16composition,kong16photo}. In this work, we develop a CNN-based aesthetic predictor and compare its performance to existing work and to human evaluation through a crowdsourcing experiment.

\vspace{5pt} \noindent \textbf{Media Content Quality and User Experience.} 
Similar to our work, several user studies in controlled lab settings have evaluated how quality affects user experience in relation to different types of media content. Gulliver et al.~\cite{gulliver06defining} found that video frame rate and network characteristics such as bandwidth and video topic impact user perception of information quality. Bouch et al. explored the importance of contextual and objective factors for media quality of service~\cite{bouch00quality}, and Ceaparu et al. found causes of user frustration in web browsing, e-mail, and word processing~\cite{ceaparu04determining}. 
In this work we explore the impact of visual aesthetic quality in online social networks. Past research has demonstrated the importance of visual aesthetics in improving user satisfaction and usability of web pages~\cite{Lavie:2004:ADP:998271.998272, DeAngeli:2006:IUA:1142405.1142446}. In the context of online advertising, researchers have found that image quality properties can impact the user experience of the ad viewed~\cite{zhou16predicting}. Aesthetically appealing preview thumbnails increase the clickthorough probability of a video~\cite{song16click}. In recent work, Schifanella et al. showed how existing features for aesthetics, embedded in topic-specific aesthetic models, can be used to surface beautiful but hard-to-find pictures and that content quality is only weakly correlated with its popularity~\cite{schifanella2015image}. We build on such work to analyze how quality production and consumption are related to the social network topology at scale.  

\vspace{5pt} \noindent \textbf{Networks and Media Diffusion.} 
Bakshy et al. examined the role of social networks in information diffusion with a large-scale field experiment where the exposure to friends' information was randomized among the target population ~\cite{bakshy12role}. They found that users who are exposed to friends' social updates are significantly more likely to spread information and do it sooner than those who are not exposed. They further examine the relative role of strong and weak ties in information propagation, showing that weak ties are more likely to be responsible for the propagation of novel information. Social exposure, assortative mixing, and temporal clustering are not the only factors that drive information diffusion and influence. Aral et al. studied the effect of homophily in explaining such evidence~\cite{aral09distinguishing}. They developed a dynamic matched sample estimation framework to distinguish influence and homophily effects in dynamic networks, and they applied it to a global instant messaging network of 27.4 million users. Stuart addressed the problem of estimating causal effects~\cite{stuart10matching} using observational data, and explained how to design matching methods that replicate a randomized experiment as closely as possible by obtaining treated and control groups with similar covariate distribution. Those type of techniques are increasingly used being used to analyze digital traces~\cite{althoff17online}; we leverage them in our work too.

%===========================================
%===========================================
\section{Dataset}  \label{sec:dataset}
%===========================================
%===========================================

Flickr is a popular photo-sharing platform on which users can upload a large number of pictures (up to 1 TB), organize them via albums or free-form textual tags, and share them with friends. Users can establish directed social links by following other users to get updates on their activity. Since its release in February 2004, the platform has gathered almost 90 million registered members who upload more than 3.5 million new images daily\footnote{This figure includes public and private photo uploads ---\url{http://bit.ly/1LjaTBT}}.

We collected a sample of the follower network composed of the nearly $40M$ public Flickr profiles that are opted-in for research studies and by all the $570M+$ following links incident to them. For each profile in the sample, we get the complete information about the \textit{photos} they upload (around 15B in total), the \textit{favorites} their photos receive from other users, and the \textit{groups} they are subscribed to. Every piece of information is annotated with timestamps that enable the reconstruction of the full temporal profile of a user's public activities. The whole data spans approximately 12 years, starting from the debut of the service in 2004 until March 2016. 

\begin{figure}[t!]
\begin{center}
\includegraphics[width=0.49\columnwidth]{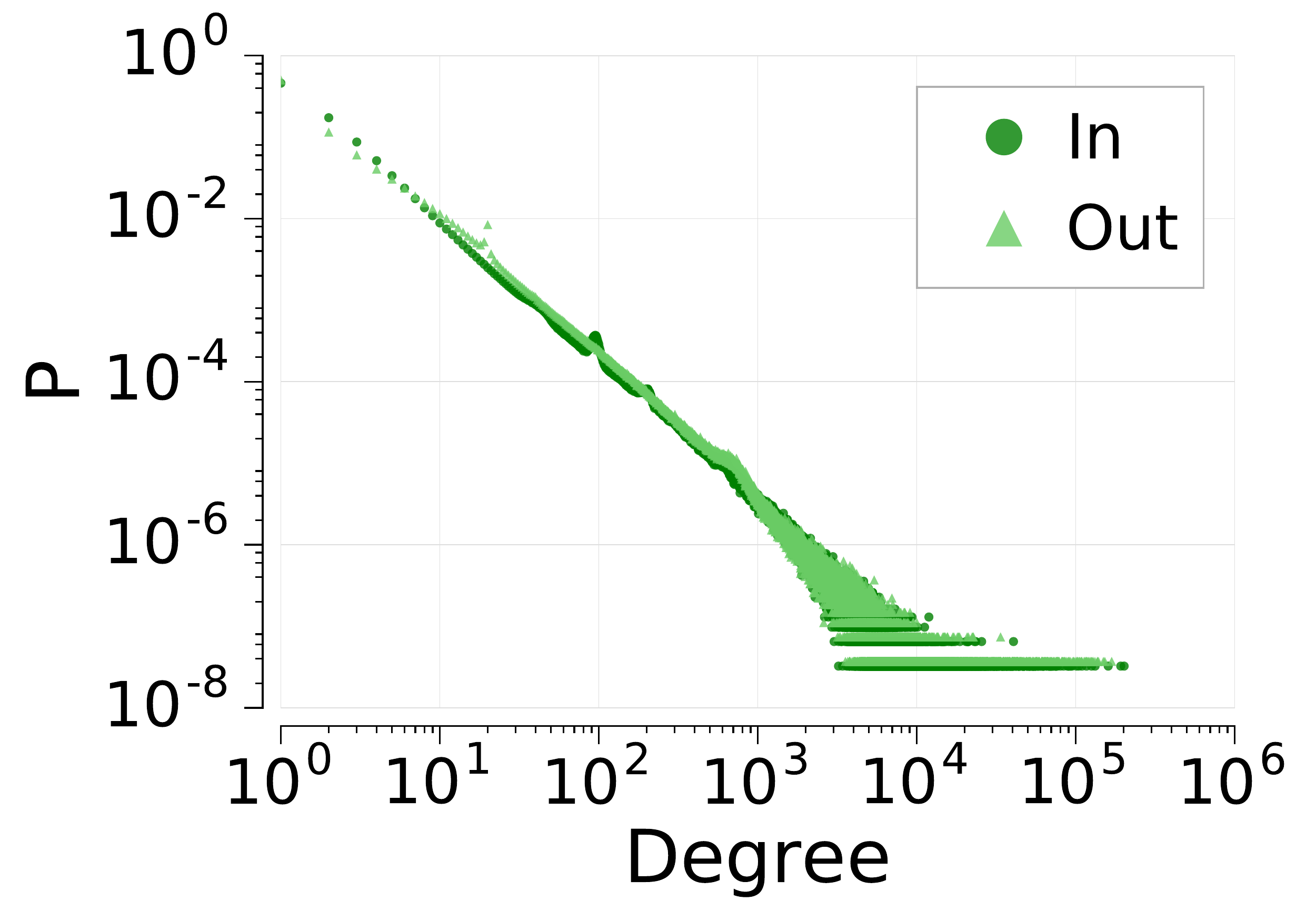}
\includegraphics[width=0.49\columnwidth]{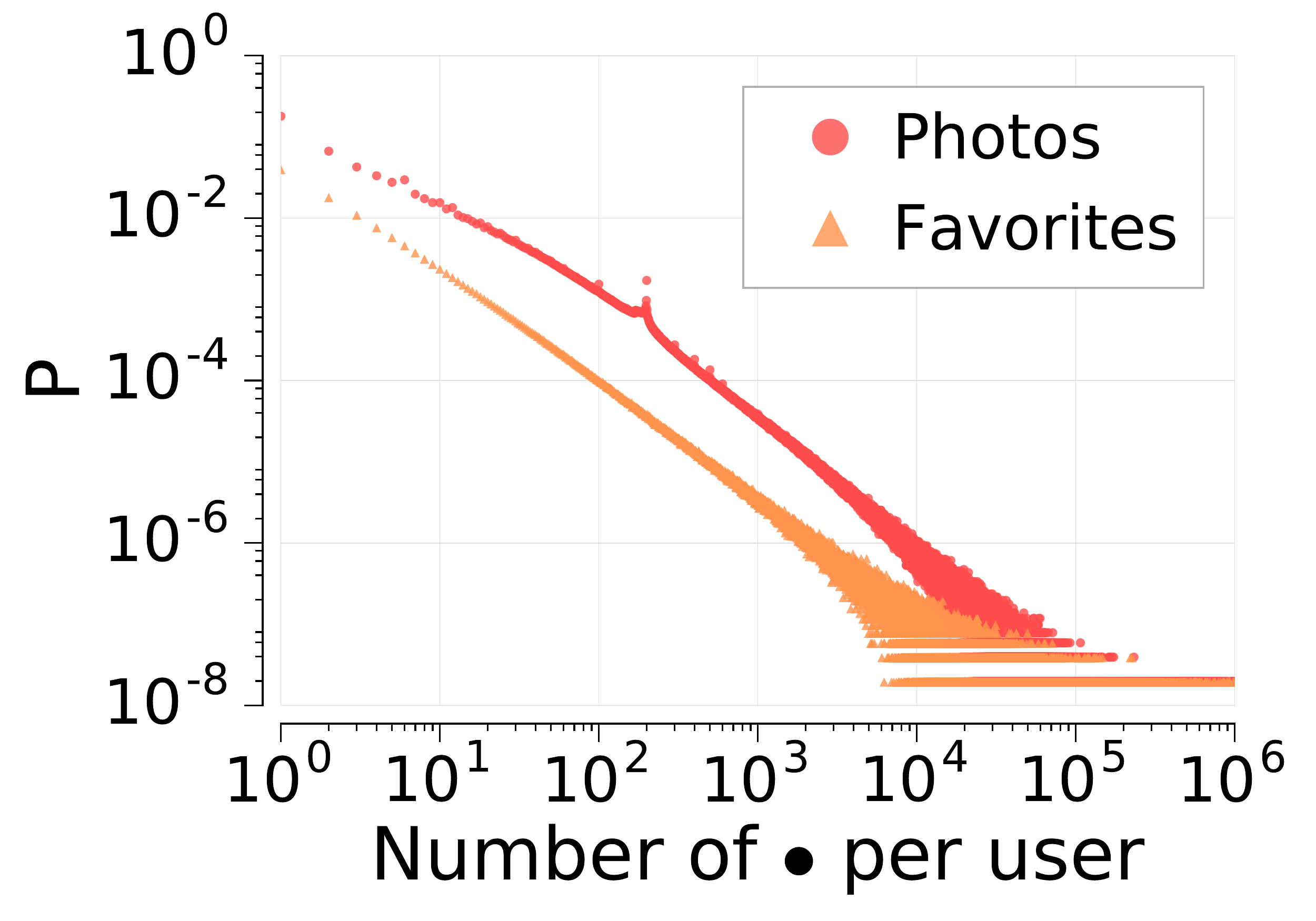}
\caption{\textit{Left:} Degree distributions $\mu_{in}=19$, $\mu_{out}=21$). \textit{Right:} Distribution of number of photos uploaded ($\mu=350$) and number of favorites received ($\mu=47$). Nearly 80\% of users receive no favorites.}
\label{fig:basic_distributions}
\vspace{-5mm}
\end{center}
\end{figure}

The distributions of the main activity and popularity indicators, along with their average values ($\mu$), are shown in Figure~\ref{fig:basic_distributions}. As expected, all distributions are broad, with values spanning several orders of magnitude.

%===========================================
%===========================================
\section{Scoring Image Beauty} \label{sec:aesthetics}
%===========================================
%===========================================

The first step towards a complete characterization of aesthetic quality in the Flickr network is to quantify beauty at the image level. To do so, we trained a deep neural network to produce a pixel-based aesthetics score. To boost performance, this network was pre-trained on a large-scale supervised image recognition task, and then the final layers were fine-tuned on our aesthetics estimation task~\cite{girshick2014rich}

\vspace{5pt}\noindent\textbf{Training vs Fine-tuning.}
Deep neural network architectures are essentially layers of artificial neurons
that progressively abstract the input data (the image pixels) into an output network response (the predicted category of the input image). In the training phase, network parameters are tuned in order to maximize metrics such as category prediction accuracy. Given the number of parameters involved in such complex architectures, effectively training neural networks is typically a long, expensive process. A common practice used to speed-up the training process is called fine-tuning, where the last layers of a trained network are modified and re-trained for a new task. In addition to making training more efficient, fine-tuning enables knowledge transfer from the original training data to the new task, improving overall performance. In our case, we start with a network designed for object detection, and then fine-tune it for the task of aesthetic scoring. This allows the aesthetic network to retain some information about the semantic nature of the objects depicted in the image, thus making the system aware of the subject depicted, which is crucial to the correct assessment of a picture's aesthetic value. As a matter of fact photographic theory~\cite{freeman2007photographer} shows that different aesthetic criteria apply to different subjects: for example, specific photographic techniques should be used when taking pictures with human subjects~\cite{hurter2007portrait}. Such observations were confirmed by  several research works in computational aesthetics~\cite{luo11content,obrador2012towards,redi15thebeauty,mai16composition}, which showed that subject-aware aesthetic scorers outperform traditional subject-agnostic aesthetic frameworks.

\vspace{5pt}\noindent\textbf{Training on Object Detection.}
We start with a network pre-trained for object detection. The architecture and training process for this network are similar to the reference model proposed by Krizhevsky et al.~\cite{krizhevsky2012imagenet}. However, we introduce a few fundamental changes. We doubled the size of the \emph{fc6} (second-last) layer from 4096 to 8192. We also used a final \emph{fc8}-layer consisting of 21841 units (instead of 1000), corresponding to the complete collection of annotated objects in the ILSVRC ImageNet dataset~\cite{ilscrvc}. We found that for the purpose of pre-training, predicting all objects was more effective than just using the standard 1000 categories typical in the ILSVRC challenges. This also allowed us to use the complete ImageNet dataset of about 14 million images.

\vspace{5pt}\noindent\textbf{Fine-Tuning on Aesthetic Scoring.}
After pre-training on the ImageNet classification task, we fine-tune the network for the aesthetic scoring task. The training set for the aesthetic quality classification task is an internal dataset created using a proprietary social metric of image quality based on Flickr's user interaction data, that has proved to correlate closely with subjective assessments of aesthetic quality. We rank all images from the YFCC100MM dataset~\cite{thomee2016yfcc100m} according to this metric and then create buckets of ``low quality'', ``median quality'', and ``high quality'' by sampling images from the bottom 10-percentile, the middle 10-percentile, and the top 5-percentile respectively. The aesthetic classification task requires the network to assign images to the right quality buckets. We then proceed to fine-tuning, replacing the final layer of the object detection network with the 3-way aesthetic quality classification task. This means that the output layer is made of 3 neurons, one for the low category, one for the medium category, and one the for high quality category. Initially, we fine-tune just the final fully connected layer; after convergence, we fine-tune the whole network.

\vspace{5pt}\noindent\textbf{Network Evaluation.}
The output layer of the network yields three scores via softmax---these correspond to the probabilities of a photo's ``low'' ($p_{LQ}$), ``medium'' ($p_{MQ}$), and ``high'' ($p_{HQ}$) quality. Each probability is the output of the corresponding neuron. Collectively, the scores correspond to the output of a softmax function evaluating the categorical probability distribution over the 3 possible outcomes: low, medium, and high. The three scores (in the range $[0,1]$) sum up to 1. In empirical evaluations, we noticed that the per-class network accuracy is higher for images in the low and high quality categories. We therefore design our continuous scoring formula by considering the output of the neurons corresponding to the low and high classes only, namely $p_{LQ}$ and $p_{HQ}$, respectively. We combine these two into a single aesthetic score by subtracting the low quality probability from the high quality probability, followed by normalization to the range [0,1]:
\begin{equation}
  s = \frac{1}{2}(p_{HQ} - p_{LQ} + 1)
\end{equation}
The network achieves a final single-crop test accuracy of 62.5\%, almost twice the accuracy of a random classifier. To further verify the performance of our approach, we compare it with state-of-the-art methods for automatic aesthetic assessment. We fine-tune the network with AVA, one of the most widely-used benchmarking datasets~\cite{murray2012ava}. Following existing work, we re-train the network for binary aesthetic classification, a simpler task compared to the 3-way decision we use, and achieve a classification accuracy of 77.6\%, thus in line with the most recent state-of-the-art on the same dataset, which stands between 75\% and 79\%, depending on the training and test setup~\cite{luo2014rapid,jin16deep,kong16photo,mai16composition}.

\vspace{5pt}\noindent\textbf{Classification vs. Regression}. We tested the possibility to predict a continuous aesthetic score using regression: we obtained a continuous aesthetic score for each sample in our training set by placing the categorical annotations on a continuous scale and normalizing in the range [0,1]; we designed the output layer to contain one single neuron predicting the aesthetic score; we trained to minimize Euclidean loss. Although this approach has been found to be effective by Kong et al.~\cite{kong16photo}, we found in empirical evaluations that this approach is less effective than our proposed methodology. As a matter of fact, our accuracy on the AVA dataset (77.6\%) is 5 points higher than the regression-based framework proposed by Kong et al.~\cite{kong16photo} (72\% for the regression based on visual data only).

%===========================================
%===========================================
\section{Crowdsourcing Beauty Assessment}\label{sec:crowdsourcing}
%===========================================
%===========================================
%
\begin{figure}[t!]
\begin{center}
\includegraphics[width=0.75\columnwidth]{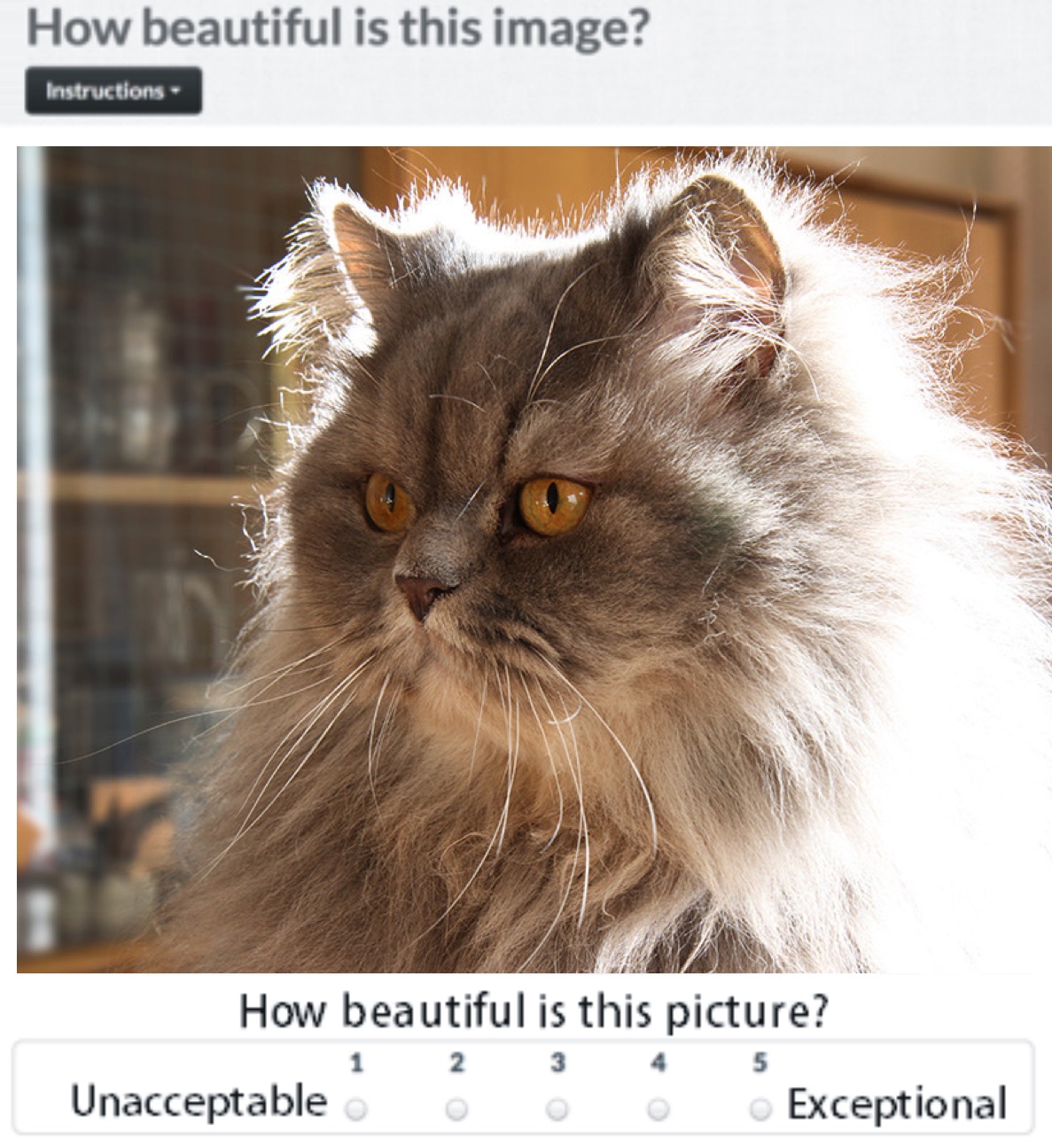}
\caption{Screenshot of the crowdflower job: instruction examples (left) and voting task (right).}
\vspace{-1.5em}
\label{fig:screenshotz}
\end{center}
\end{figure}
\begin{figure*}[t!]
\begin{center}
\includegraphics[width=0.19\textwidth]{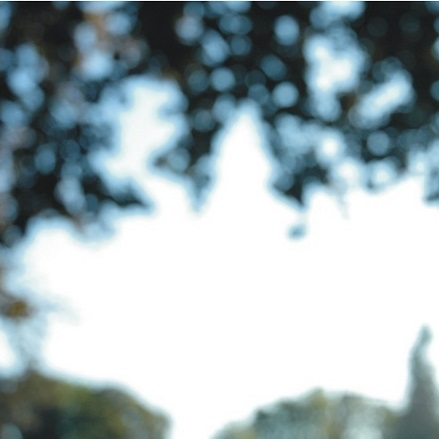}
\includegraphics[width=0.19\textwidth]{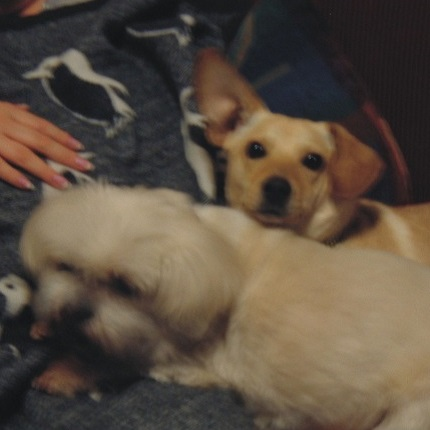}
\includegraphics[width=0.19\textwidth]{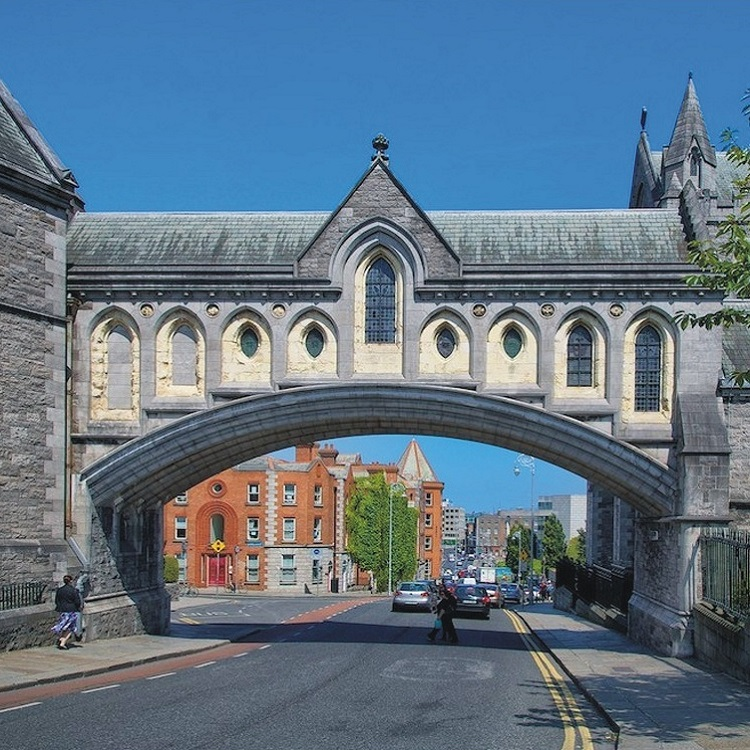}
\includegraphics[width=0.19\textwidth]{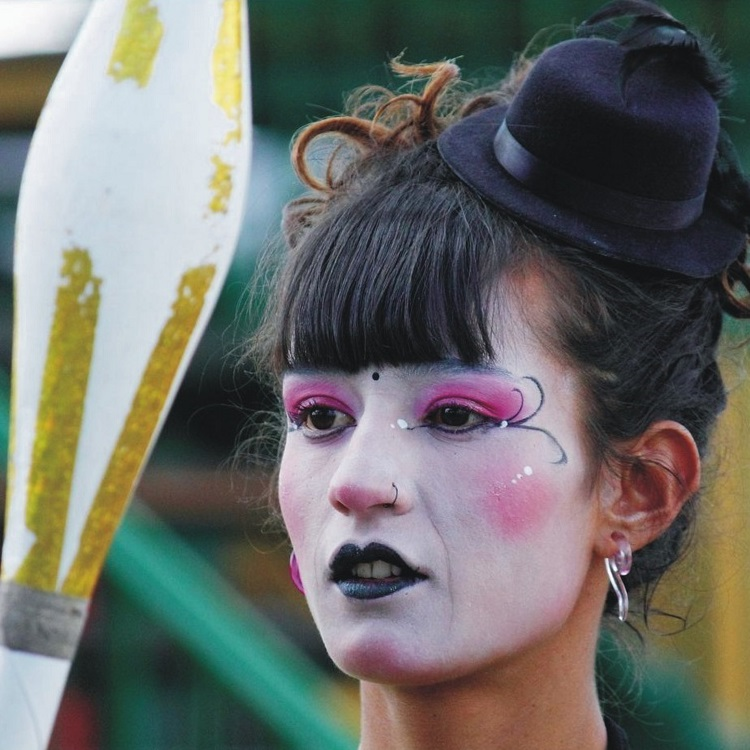}
\includegraphics[width=0.19\textwidth]{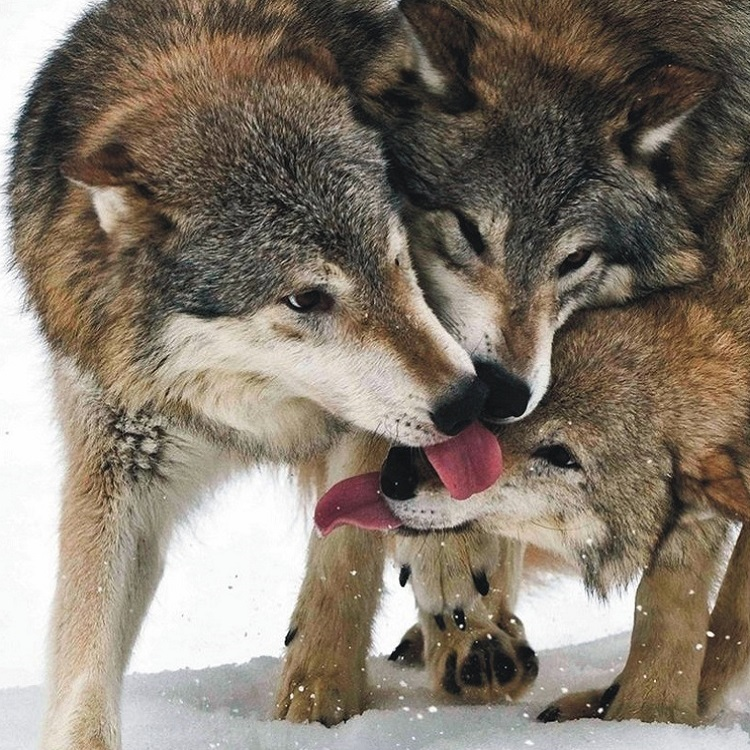}
\caption{Examples of images ranging from beauty score 1 (leftmost) to score 5 (rightmost). These and other examples were provided to crowdworkers for the sake of training.}
\vspace{-1.5em}
\label{fig:screenshotz_grades}
\end{center}
\end{figure*}

In addition to the standard performance test on benchmarking datasets, we further evaluate the effectiveness of the aesthetic network with a crowdsourcing experiment. We ask people to evaluate pictures in terms of their beauty, and then compare the human judgments to the aesthetic score predicted by our framework. To design our experiment, we draw inspiration from the image beauty assessment crowdsourcing experiments conducted by Schifanella et al.~\cite{schifanella2015image}.

Crowdsourcing tasks are complex and can be influenced by unpredictable human factors~\cite{mason14conducting}. Modern crowdsourcing platforms offer control mechanisms to tune the annotation process and enable the best conditions to get high-quality judgments. To annotate the beauty of our images, we use CrowdFlower\footnote{\url{http://www.crowdflower.com/}}, a popular crowdsourcing platform that distributes small \textit{tasks} to online \textit{contributors} in an assembly line fashion.

\vspace{5pt}\noindent\textbf{Data selection}.
To help the contributor to assess the image beauty more reliably, we build a photo collection that represents the full popularity spectrum, thus ensuring a diverse range of aesthetic values. To do so, we identify three popularity buckets obtained by logarithmic binning over the range of number of favorites $f$ received. We refer to them as \textit{tail} ($f\leq5$), \textit{torso ($5<f\leq45$)}, and \textit{head} ($f>45$) to identify the characteristic segments of the broad distribution. From the validation set used to evaluate the aesthetic network, we randomly sample 1000 images from each bucket. Images from such diverse popularity levels are also likely to take a wide range of aesthetic values, thus ensuring aesthetic diversity  in our corpus, typically very important for the crowdsourcing of reliable beauty judgments~\cite{redi2013crowdsourcing}.

\vspace{5pt}\noindent\textbf{Crowdsourcing task setup}.
The task consists in looking at a number of images and evaluating their aesthetic quality. At the top of the page we report a short description of the task and we ask to answer the question \textit{``How beautiful is this image?''} (Figure~\ref{fig:screenshotz}). The contributor is invited to judge the intrinsic beauty of the image and \textit{not the appeal of its subject}; for example, artistic pictures that capture non-conventionally beautiful subjects (e.g., a spider), should be considered beautiful. Out of all the possible rating scales commonly used in crowdsourcing~\cite{fleet14interestingness}, it has been shown that the 5-point \textit{Absolute Category Rating} (ACR) scale is good way to collect aesthetic preferences~\cite{redi2014beauty}. We therefore ask contributors to express their judgments by selecting one out of 5 aesthetic categories from \textit{``Unacceptable''} to \textit{``Exceptional''}. To guide the contributor in its choice, two example images for each grade are shown (Figure~\ref{fig:screenshotz_grades}). Examples are Flickr images that have been unanimously judged by three independent annotators to be clear representative instances of that beauty grade. Below the examples, the page contains 5 randomly selected images to be rated. The images in each page are randomly selected and displayed in an approximate equally-large size to minimize any skew in the perception of image quality~\cite{chu13size,fleet14interestingness}.

\vspace{5pt}\noindent\textbf{Quality control}. To maximize the quality of human judgments, we apply several controls on the contributors' input. First, we open the task only to Crowdflower contributors with an \textit{``excellent''} track record on the platform (responsible for the $7\%$ of monthly CrowdFlower judgments). We also limit the task to contributors from specific countries\footnote{Australia, Austria, Belgium, Denmark, Finland, France, Germany, Ireland, Italy, Netherlands, Poland, Spain, Sweden, United Kingdom, United States}, to ensure higher cultural homogeneity in the assessment of image beauty~\cite{hagen78perception,russell94universal,miyamoto06culture,dewar07photographic,yanai09mining}.
Second, we cap the contributions of each worker to a maximum of 500 judgments to prevent potential biases introduced by the predominance of a small group of active workers. Last, we discard all the judgments of contributors who did not annotate correctly at least 6 out of 8 \textit{Test Images} that are presented to them in an initial \textit{Quiz} page and randomly throughout the task, disguised as normal units. Similar to the examples, Test Images are Flickr pictures that have been unanimously judged by three annotators to be clear representative instances of a beauty score. 

\vspace{5pt}\noindent\textbf{Agreement}.
Each photo receives at least 5 judgments by as many independent contributors. Despite aesthetics assessments having a strong subjective component, we register a good level of agreement between annotators, in line with previous work on image beauty~\cite{schifanella2015image}. The average percentage of matching annotations over 5 judgments is $73\%$. When judgments do not match exactly, they usually cluster around two consecutive scores; the average standard deviation around the average score is $0.45$, less than half point. In alternative to matching, we also compute Cronbach's $\alpha$, a widely-adopted metric to assess inter-rater agreement on aesthetics tasks~\cite{redi2014beauty}. The Cronbach's coefficient is $0.77$, a value that falls in a range that is commonly considered a \textit{Good} level of inter-rater consistency~\cite{bland1997statistics}.

\begin{figure}
\begin{center}
\includegraphics[width=0.70\columnwidth]{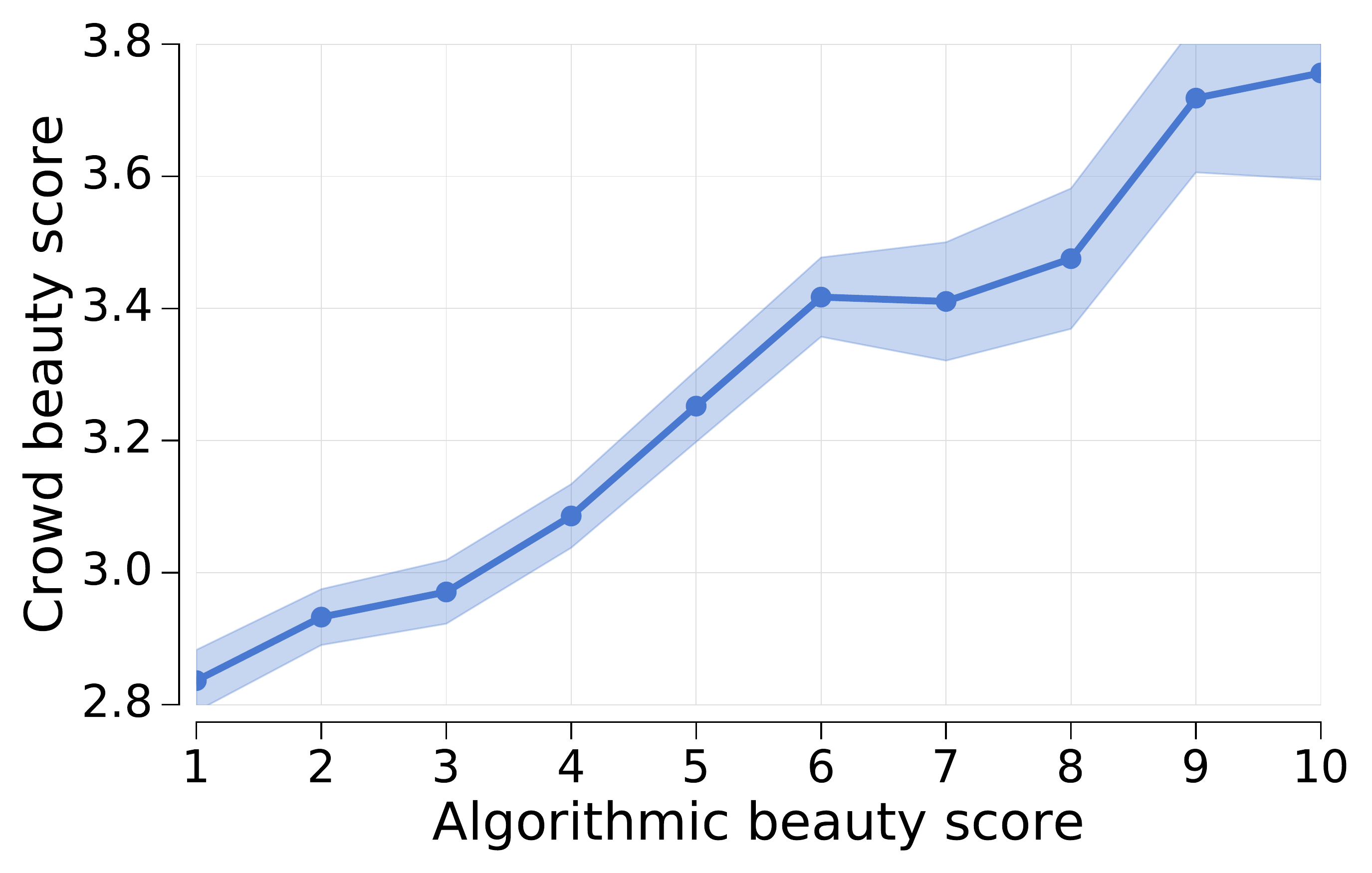}
\caption{Average beauty as assessed by crowdworkers against the algorithmic beauty from our deep learning model. Spearman correlation $\rho=0.48$. 95\% confidence interval is shown.}
\label{fig:beauty_vs_crowd}
\vspace{-3mm}
\end{center}
\end{figure}

\vspace{5pt}\noindent\textbf{Results}.
Having collected reliable annotations on 3,000 validation images, we test the aesthetic network predictions relative to the ground truth as follows. We are interested in a predicted score that, regardless of its range or distribution, preserves the ranking of the original beauty scores assigned by human annotators. To check that, we compute the Spearman rank correlation coefficient $\rho$ between the predicted score and the crowdsourced score. We find a high correlation $\rho=0.48$ (with $p<0.01$), which suggests that our automatic aesthetic scoring method is an effective proxy of human aesthetic judgment. To further dig into this intuition, we partition the validated images into 10 equally-spaced intervals of predicted aesthetic score (i.e., $[0,0.1], ...,[0.9,1]$). We then compute the average crowd-sourced beauty score for all images in each bucket. Figure~\ref{fig:beauty_vs_crowd} shows that the average crowdsourced score linearly increases with the predicted beauty decile, further confirming that our aesthetic framework performs comparably to human evaluation on this task.

Additionally, we test the level of agreement of the algorithmic beauty prediction with the judgment of human labeler using the state of the art approach proposed by Ye et al.~\cite{ye2017probabilistic}. Their evaluation method, inspired by the work on consensus methods by Dawid and Skene~\cite{dawid1979maximum}, has been used to assess the robustness of crowdsourced affective data and can be used to estimate how much machine-generated labels can accurately mimic the human judgments. We apply the method to the human-generated beauty scores and found an average reliability score of $\overline{\tau}=0.71$, with peaks of $\tau_{max}=0.95$, much higher than the reliability of a random annotator $\tau_{rand}=0.22$ (to obtain this number, we added to the pool of annotators a fake annotator giving random judgments). Next, we re-scale the continuous scores predicted by the aesthetic network over a discrete 5-point scale, in order to make machine predictions comparable to human labels. We add the scaled predictions to the previous list of judgments by treating the machine-generated scores as the output of an additional annotator. We re-calculate reliability of all annotators, including the machine: we find that the reliability of the machine judgments stands at $0.77$, in line with the average reliability score.

%===========================================
%===========================================
\section{Network Effects} \label{sec:network}
%===========================================
%===========================================

While previous work has studied beauty at the picture level, our large-scale rating of image beauty further enables us to analyze the how beauty is produced over a large social network. In the following,  we will characterize the beauty $\bar{b}(i)$ of a user $i$ as the average beauty of all of $i$'s public photos. We will refer to this score as \textit{user beauty} or \textit{user quality}, for brevity. When time is relevat to the analysis, we will use $b^t(i)$ to denote the average beauty of pictures posted by user $i$ during week $t$ and  $\bar{b}^t(i)$ to denote $i$'s photos average beauty \textit{until} week $t$. Although summarizing the quality production of a user with a single indicator is limiting, it helps to simplify the analysis that follows. In future work we plan to consider more complex quality profiles that include, for example, the variance of photo quality.

\begin{figure}[t!]
\begin{center}
\includegraphics[width=0.49\columnwidth]{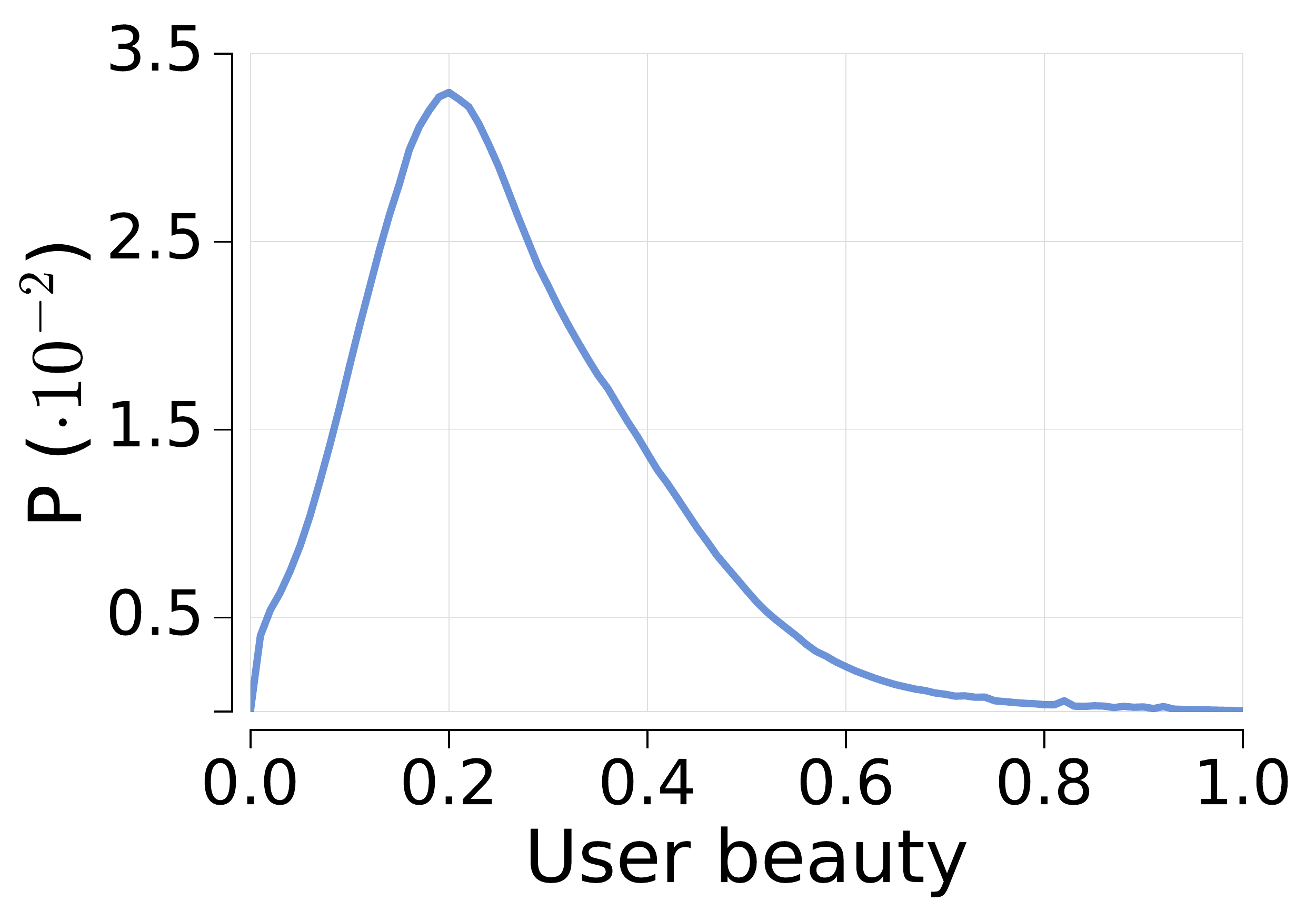}
\includegraphics[width=0.49\columnwidth]{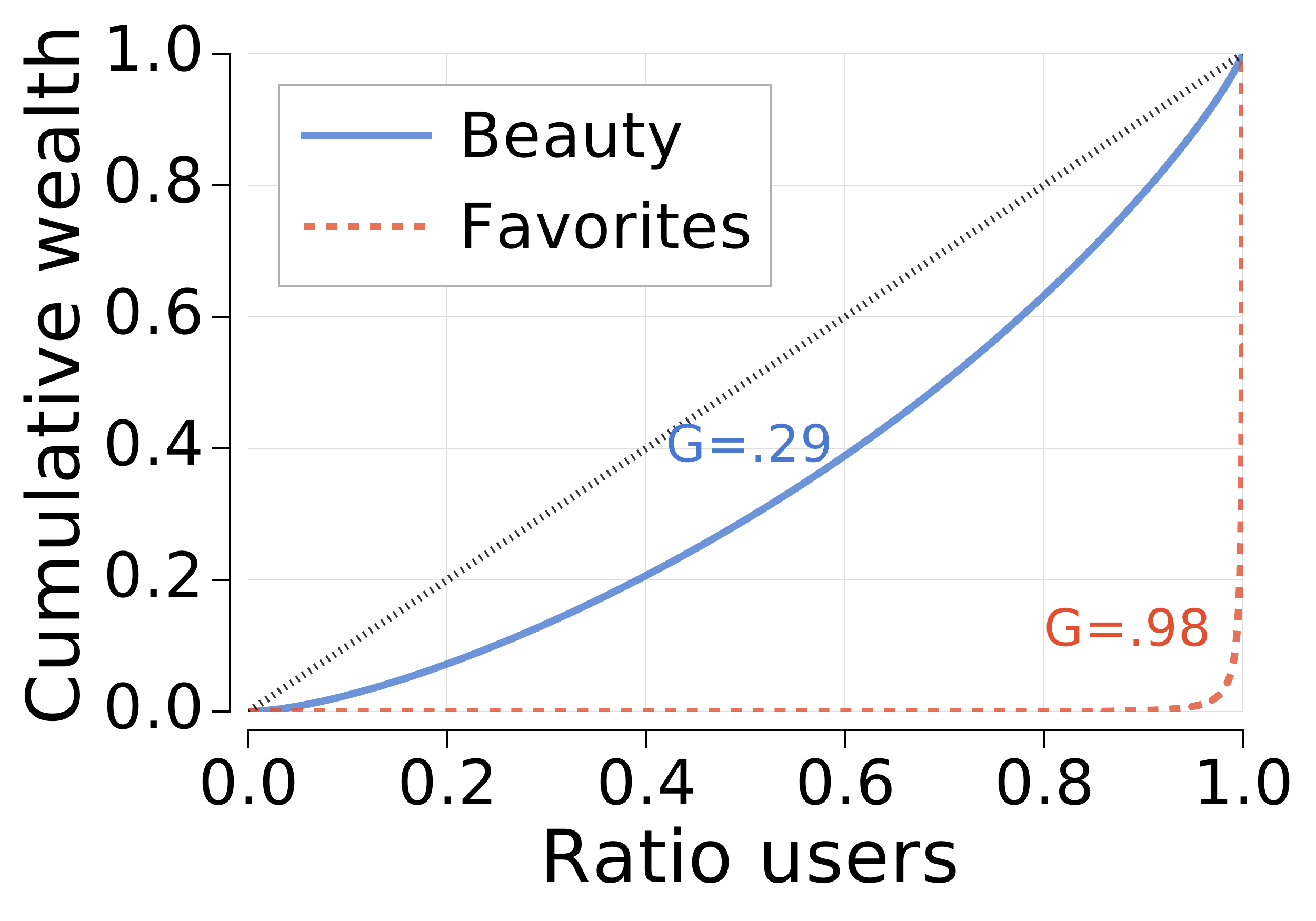}
\caption{Distribution of beauty scores; $\mu=0.26$ (left). Inequality of resource distribution (average beauty and average favorites) across users visualized with the Lorenz curve. Gini coefficients: $G_{favs}=0.98$, $G_{beauty}=0.29$ (right)} %Photos ($G=.92$,$H=0.79$), % Beauty*numphotos $G=0.91$,$H=0.78$
\label{fig:beauty_distribution_inequality}
\vspace{-5mm}
\end{center}
\end{figure}

Unlike the heavy-tailed distributions of activity and popularity indicators (Figure~\ref{fig:basic_distributions}), the user beauty is bell-shaped distributed, with a slightly heavier right tail (Figure~\ref{fig:beauty_distribution_inequality}, left). This leads to a mismatch between the ability to produce high-quality content and the social attention received by the community. As a result, we observe a more marked inequality in the distribution of the average number of favorites per photos across users than in the distribution of average user quality (as measured by the Gini index, Figure~\ref{fig:beauty_distribution_inequality}, right). This finding is in line with previous work on a smaller data sample~\cite{schifanella2015image} showing that high-quality Flickr pictures are distributed across different ranges of popularity.

In other words, this heavy imbalance reveals that a large number of users who post high-quality photos receive very little social attention. Next, we map the average user beauty on the Flickr follower network to further investigate the unexplored relationship between user beauty and social connectivity patterns. In particular, we are interested to shed light on two unexplored matters: \textit{i)} how the quality is distributed over the network ($\S$\ref{sec:network:mixing}) and \textit{ii)} the causal impact that the quality users are exposed to has on their own activity and engagement ($\S$\ref{sec:network:causal}). These issues are crucial to managers of online communities, who aim to provide all users with high-quality content and retain them as long as possible. However, those could not be addressed in the past due to the scarcity of large-scale data suitable for such analysis and the lack of reliable and efficient tools to measure content quality.

\subsection{Distribution of quality over the social network}  \label{sec:network:mixing}

Different activity indicators of social media users tend to be correlated. This has been verified in multiple social media platforms, including Flickr, on a wide range of indicators, especially in relation to nodal degree~\cite{mislove07measurement,schifanella10folks}. We are interested in verifying whether the level of user quality is correlated to social connectivity or other activity indicators.

\vspace{5pt}\noindent\textbf{Q1: Is quality correlated with social connectivity?}
We compute the Spearman rank correlation $\rho$ between user beauty and nodal degree. We find \textbf{A1:} \textit{a small but positive correlation $\rho$ between user beauty, indegree ($\rho=.22$), and outdegree ($\rho=.24$)} (Figure~\ref{fig:correlations_vs_beauty}). Beauty is also weakly associated with the average number of favorites received by a user ($\rho=.17$) and exhibits a slightly negative correlation with the number of photos posted ($\rho=-.03$), confirming that neither content popularity nor volume of contributions are strong determinants of quality.

The association between quality and connectivity can have higher-order effects. In online social networks, as in offline social environments, people lack global knowledge of the overall population's characteristics, since their view of the external world is mediated by their direct social connections. This local constraint might lead to an over-representation of some rare population attributes in local contexts. This phenomenon has been observed in the form of the so-called \textit{friendship paradox}~\cite{feld91your,hodas13icwsm}, a statistical property of networks with broad degree distributions for which on average people have fewer friends than their own friends. The paradox has been recently extended by the concept of \textit{majority illusion}~\cite{lerman16majority}, which states that in a social network with broad degree distribution and binary node attributes there is a systematic biased local perception that the majority of people (50\% or more) possess that attribute even if it is globally rare. As an illustrative example, in a network where people drinking alcohol are a small minority, the local perception of most nodes can be that the majority of people are drinkers just because drinkers happen to be connected with many more neighbors than the average.

In our context, we are interested in measuring the presence of any skew in the local perception of the quality of user-generated content. So we ask:

\begin{figure}[t!]
\begin{center}
\includegraphics[width=0.48\columnwidth]{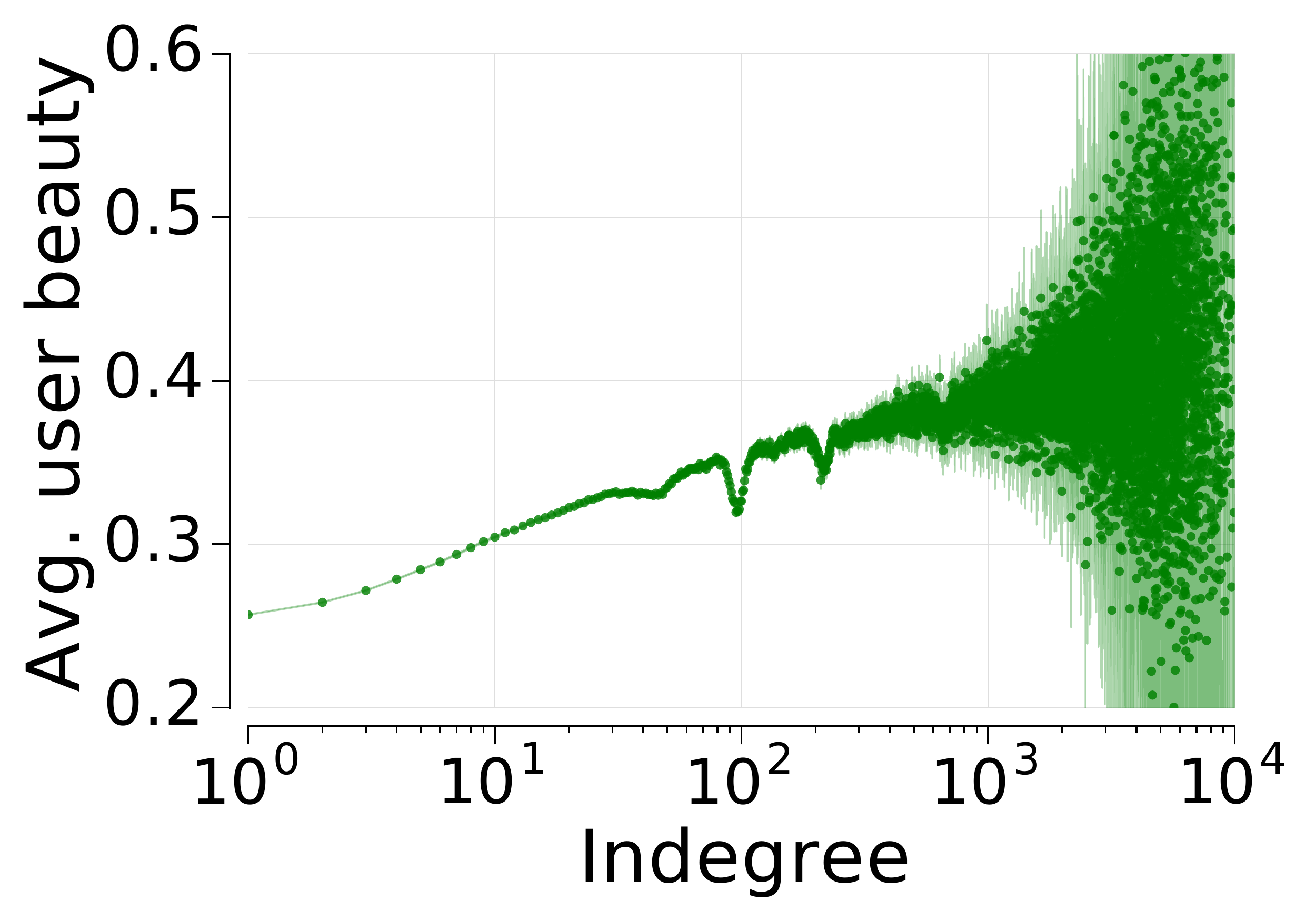}
\includegraphics[width=0.48\columnwidth]{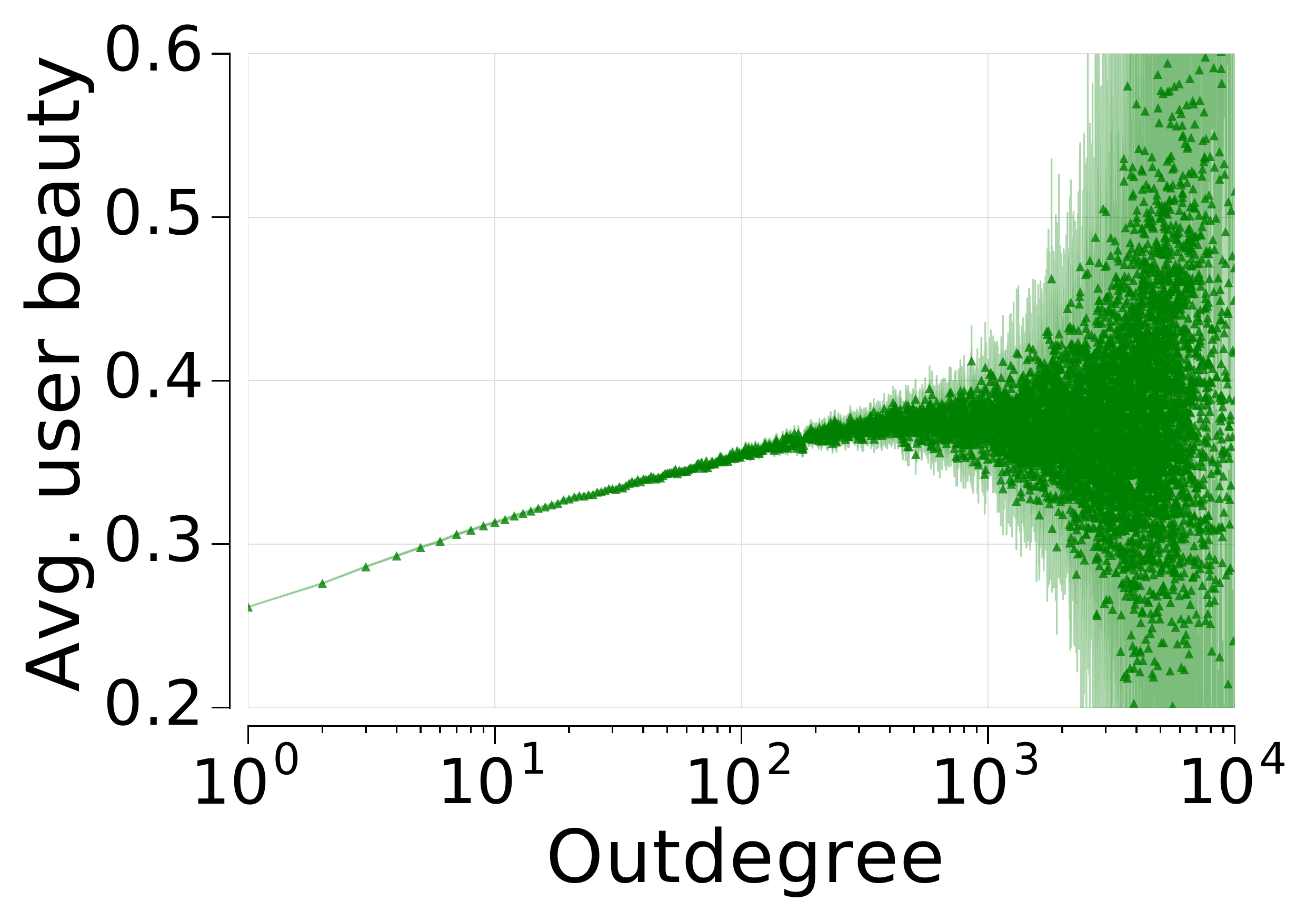}
\caption{Average user beauty for users with fixed indegree (left) and outdegree (right). The positive slopes (Spearman correlations $\rho=.22$ and $\rho=.24$, respectively) indicate that users who are more connected tend to produce higher-quality content. 95\% confidence intervals are shown.}
\label{fig:correlations_vs_beauty}
\vspace{-5mm}
\end{center}
\end{figure}

\vspace{5pt}\noindent\textbf{Q2: Does the correlation between connectivity and quality creates a majority-illusion effect on user beauty?}

\noindent To estimate the presence of any local perception skew, we calculate the proportion of users in a node's neighborhood whose quality is above the global average quality of users in the network ($\mu=0.26$, as per Figure~\ref{fig:beauty_distribution_inequality} (left)), and compare it with the actual portion of users in the overall population with beauty above the global average\footnote{\small{Because the overall distribution of quality has a shape that is close to normal, the results do not change considerably when using the median instead.}}. We find that \textbf{A2:} \textit{the majority illusion holds in our data sample. Overall, $43\%$ of the users typically produce content with above-average quality; however, $65\%$ of the population has more than $43\%$ of their friends with above-average quality.} The phenomenon is very strong for the nearly $20\%$ of users who have more than $86\%$ of their neighbors falling into this category (double or more than what is expected). Nevertheless, the majority illusion does not imply that people preferentially connect to very talented users. Next, we investigate the relationship between the beauty levels of connected individuals.

\vspace{5pt}\noindent\textbf{Q3: Are social connections established between users with similar beauty?}

\begin{figure}[t!]
\begin{center}
\includegraphics[width=0.80\columnwidth]{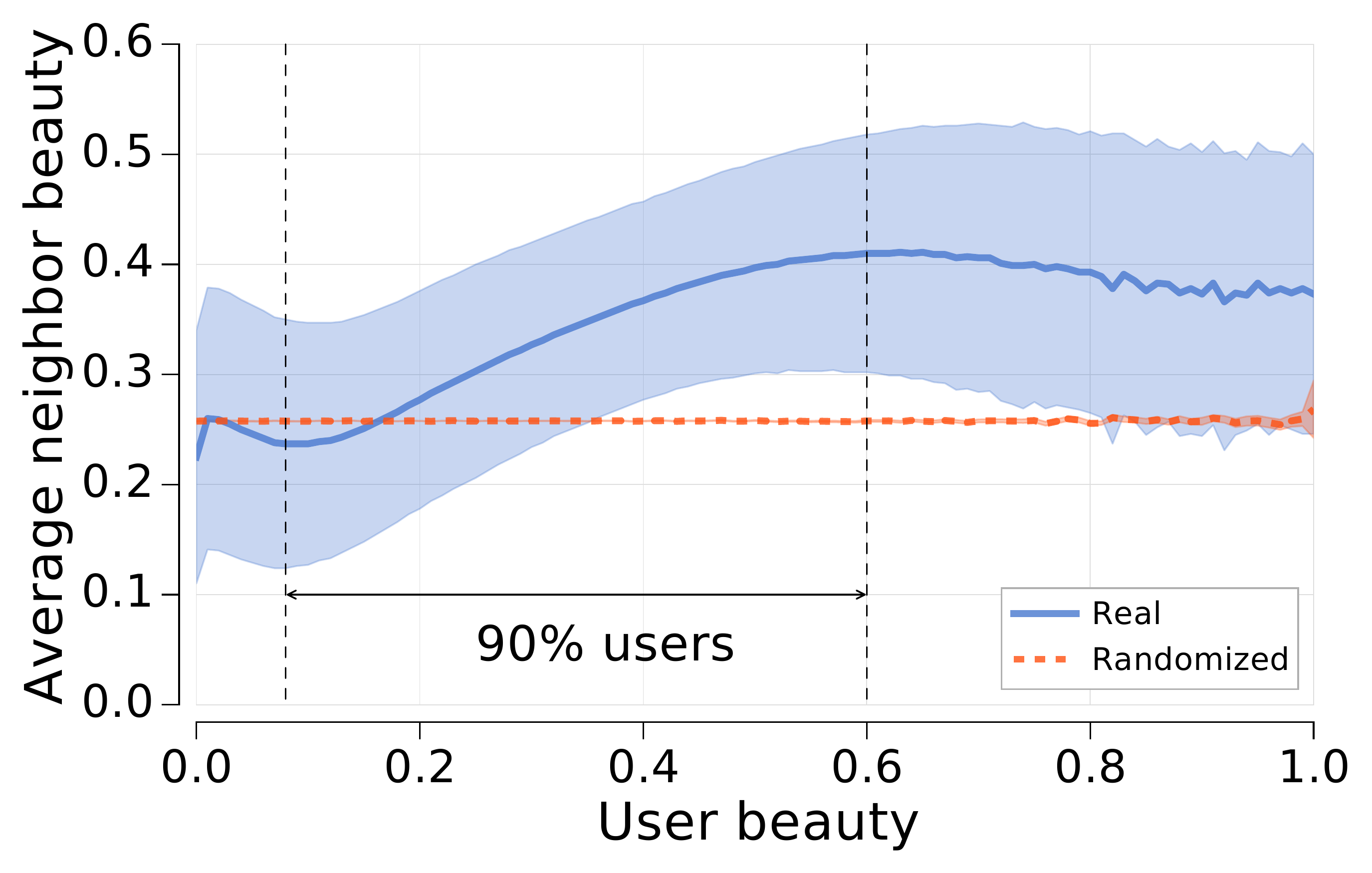}
\caption{Correlation spectrum of user beauty on the Flickr follower network. The highlighted interval on the beauty axis includes the user beauty values of $90\%$ of the population. Variance is shown. The correlation spectrum for a null-model with randomly reshuffles user beauty scores is reported for the sake of comparison.}
\label{fig:assortativity}
\vspace{-5mm}
\end{center}
\end{figure}

\noindent A typical pattern found in several ecological and social networks is \textit{assortative mixing}, namely the high likelihood of nodes to be connected to other nodes with similar properties. This propensity is gauged with the \textit{correlation spectrum}~\cite{barrat08dynamical}, a measure that puts in relation all the nodes that have a fixed value $k$ of a target indicator with the average value of the same indicator of their neighbors. By setting user beauty as the target indicator, we measure the correlation spectrum by computing the average neighbor beauty of all those users with a fixed user beauty $\bar{b}=k$, for all possible values of user beauty
\begin{equation}
b_{nn}(k) = \frac{1}{|\{i: \bar{b}(i) = k\}|} \cdot \sum_{i: \bar{b}(i) = k}{\frac{\sum_{j \in \Gamma_{out}(i)}{\bar{b}(j)}}{|\Gamma_{out}(i)|}}
\end{equation}
where $\bar{b}(i)$ is user $i$'s beauty and $\Gamma_{out}(i)$ are $i$'s out-neighbors. Figure~\ref{fig:assortativity} shows the trend of $b_{nn}$ for all possible values of $k \in [0,1]$, obtained by partitioning the beauty range into 100 equally-sized bins. The positive slope of the curve (Spearman correlation $\rho=0.48$) reveals an assortative trend, which indicates that \textbf{A3}: \textit{users tend to be linked to accounts that publish photos with similar quality as their own}. The trend is particularly clear for users with beauty in the range [0.08,0.6], which corresponds to $90\%$ of our sample's population. To tell apart real any assortativity trend from statistical artifacts, we need to compare the results obtained on the real data with a suitable \textit{null model}. When using a null model that randomly reshuffles the beauty values between all users, keeping unchanged their social connections, the trend is lost.

\vspace{3pt}
In summary, we have found that user quality correlates with individual connectivity, which in turn leads to a majority illusion phenomenon, where high-quality users are more visible than lower quality ones. Also, beauty is an assortative property, with user being preferentially connected to others with similar quality.

\subsection{Network effects on user retention and quality production}\label{sec:network:causal}

\subsubsection{Quality, network, and engagement} \label{sec:network:engagement}

The assortative mixing of quality in the social network could be ascribed mainly to homophily or influence~\cite{aiello12tweb}. On one hand, users might preferentially connect to accounts that publish pictures with a similar quality to their own. This would seem natural in a platform like Flickr that hosts a heterogeneous user-base: semi-professional photographers might be interested in following users who are well-versed in the use of photographic techniques, whereas casual users might be following each other mostly for social reasons, unconcerned about aesthetic photo quality. On the other hand, pairs of users might be imbalanced in terms of their quality at the time they connect and close their quality gap later on, over time. For example, amateur photographers could follow professionals and learn new skills from them, thus improving the quality of their pictures.

The interplay between homophily, influence, and other factors leading to assortative mixing has been the subject of a number of studies~\cite{anagnostopoulos08influence,crandall08feedback} that explored these phenomena on a wide range of user attributes (e.g., demographics, topical preferences). However, despite its crucial role in growing and maintaining user engagement~\cite{Dobrian:2011:UIV:2018436.2018478}, content quality has never been investigated in relation to such network properties. We aim to shed light on this relationship by answering two research questions that help explain the assortative trend we found.

\vspace{5pt}\noindent\textbf{Q4: Is the user beauty affected by the content produced by their social neighbors?}
The quality of content produced by users might be affected by the quality of the content that their social contacts produce. In particular, we hypothesize that, on average, \textit{the user beauty increases as an effect of the creation of a new social connection with a higher-beauty user.}

\vspace{5pt}\noindent\textbf{Q5: Does a heavy quality imbalance between connected individuals affect their social engagement?} We hypothesize that, on average, \textit{heavy imbalance between the user ebauty and the average beauty of its neighbors leads to a drop in engagement.} This intuition is backed by one of the core principles of the Social Exchange Theory~\cite{blau1964exchange}, which states that reciprocity is necessary to maintain a stable social relationship. When reciprocity fails consistently, at least one of the parts is likely to withdraw. In online social platforms, users join with specific expectations; when those are not met, the likelihood of abandonment is expected to rise. Specifically in the context of Flickr, talented photographers won't feel their efforts being reciprocated if the quality of all other contributors' content is mediocre, whereas casual photographers might feel overwhelmed if mostly surrounded by professionals and will more likely regress to a lurking state or even unsubscribe.

\subsubsection{Matching experiments for causal inference} \label{sec:network:matching}

To answer the two questions above, we set up matching experiments aimed at inferring causality from the observational data. In natural experiments, estimating the statistical effect of a treatment on a population can be done through \textit{randomization}. Provided that the population is sufficiently large, randomly allocating individuals across the \textit{treatment} and \textit{control} groups cancels the potential biases by equalising all the observable factors as well as unobserved variables that have not been explicitly accounted for. Without the possibility to run controlled experiments over the Flickr user-base, we need to infer causality from observational data. That is a much harder task~\cite{aral09distinguishing,shalizi11homophily} because the benefit of randomization is lost, as the set of individuals who received the treatment is often pre-determined.

Matching experiments provide a way to reliably estimate the statistical effect of a treatment on a dependent variable from longitudinal data. The key intuition is to match the treated group $G_t$ with a control group $G_c$ whose members did not receive the treatment and are statistically indistinguishable (i.e., only marginally different) from the treated group on all observable covariates.

There are several ways to perform matching~\cite{rosenbaum02observational,stuart10matching,olteanu17distilling} and to measure the equivalence between treatment and control groups. Here we borrow a framework introduced by Rubin~\cite{rubin01using} and later summarized by Stuart~\cite{stuart10matching}, which has been successfully used in other observational studies aimed at infer causality~\cite{althoff17online}. This framework assumes that $G_t$ and $G_c$ are somehow formed and provides a function to check their statistical equivalence. The two groups are said to be \textit{balanced} on a covariate $X$ when the covariate's \textit{standardized bias} $SB$, namely the difference of its mean values ($\bar{X}$) in the two groups divided by the standard deviation ($\sigma$) in the treated group, is under a given threshold commonly set to $0.25$. Formally:
\begin{equation}
SB_X(G_t,G_c) = \frac{\bar{X_t} - \bar{X_c}}{\sigma(X_t)} \leq 0.25
\end{equation}
The groups are overall \textit{balanced} ---and therefore indistinguishable, from a statistical point of view--- only if they are balanced on \textit{all} their covariates.

\begin{table}[t]
\begin{center}
\begin{tabular}{c|ll}
\textbf{Category} & \textbf{Covariate} & \textbf{SB}\\
\hline
\textit{User} & Indegree & $+0.16$ \\
& Outdegree & $+0.21$\\ 
& Number of photos uploaded & $-0.16$ \\
& Number of group memberships & $+0.21$\\ 
& Number of favorites given & $+0.18$ \\ 
& Number of favorites received & $+0.17$\\ 
& Average photo beauty & $-0.07$\\ 
& Weeks elapsed from join date & $+0.19$\\
\hline
\textit{Neighbors} & Number of photos uploaded & $+0.18$\\
& Average photo beauty & $+0.22$ \\ 
\hline
\textit{New neighbors} & Number of photos uploaded & $+0.18$ \\
\end{tabular}
\caption{Covariates accounted for in the matching experiments. The variables considered are measured for three types of users: \textit{i)} the users who creates new links, \textit{ii)} their neighbors before the action of link creation, \textit{iii)} their new neighbors. All the measurements are taken in the week of link creation. The standardized bias values (SB) for the first matching experiment are reported.}
\label{tab:covariates}
\vspace{-5mm}
\end{center}s
\end{table}

\vspace{5pt} \noindent \textbf{Algorithm to balance treatment and control groups.} Given a treatment group $G_t$, we set a greedy iterative procedure to select a corresponding balanced control group $G_c$. At step 1, a candidate control group $G^1_c$ such that $|G_c| \gg |G_t|$ is selected from the set of non-treated units. At step $n$, the standardized bias $SB(G_t,G^n_c)$ is computed for every covariate. For all the covariates that do not satisfy the balance constraint, we remove from the control group the elements that most contribute to the mismatch. Specifically, we cut off the $1\%$ of experimental units with the highest values of the covariate, when $SB$ is negative, or with the lowest values, when $SB$ is positive. At each iteration, further pruning could be required on different sets of covariates. The algorithm stops when the condition $SB(G_t,G_c) \leq 0.25$ is satisfied by all the variables. The procedure does not have a theoretical guarantee to stop before pruning out all the elements of $G_c$, in which case the algorithm should be restarted with a different seed control group. In our experiments we always observe convergence before $|G_c| < |G_t|$.

\vspace{5pt}\noindent Next, we describe how this framework is instantiated on our Flickr data. For these experiments, we have considered only users with at least 10 outgoing social links (i.e., followees) and who have uploaded photos in at least 12 distinct weeks (which implies they all have at least 12 photos each). This filtering step yielded a subset of 2.7M users.

\subsubsection{The effects of neighbors' beauty} \label{sec:network:matching_effect}

\vspace{3pt}\noindent\textbf{Beauty inspires beauty}. To answer question \textbf{Q4}, we use link creations as events to split users between treatment and control groups. We include in the treatment group users who have created a social link towards accounts with higher quality than their own and compare them with a control group whose members have connected to users with equal or lower quality.

\begin{figure}[t!]
\begin{center}
\includegraphics[width=0.99\columnwidth]{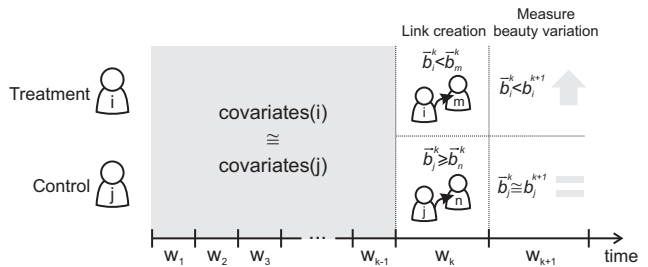}
\caption{Simplified example of matching experiment setup, with just one user in each group. The two users are statistically equivalent with respect to all the considered covariates measured at week $k-1$. At week $k$, user $i$ creates a link towards user $m$, whose photos posted until week $k$ have higher beauty than $i$'s ($\bar{b}^k(i) < \bar{b}^k(m)$). User $j$ instead, creates a new link towards user $n$, whose beauty is not higher than his ($\bar{b}^k(j) \geq \bar{b}^k(n)$). User $i$ is the treatment user, user $j$ is the control one. At week $k+1$ both users will post new photos; the hypothesis is that the $i$'s new photos will have higher quality than $i$'s previous quality ($b^{k+1}(i) > \bar{b}^k(i)$), whereas no statistically significant variation will occur to $j$'s beauty.}
\label{fig:scheme}
\vspace{-5mm}
\end{center}
\end{figure}

Operationally, we partition the timeline of events in our data into discrete slots of one week each. For each week $w$, we iterate over the set of users $U_w$ who have been active during that week and have been active for at least 12 non-consecutive weeks before it. All users who added at least one link towards higher-quality users on that week are added to the control group $G_t$. Among the remaining users in $U_w$, we add to $G_c$ those who created any number of links during that week. Each element in the two groups is described with a vector of \textit{covariates} that accounts for all the main aspects related to the popularity, activity, age, and quality of the users and to the quality and activity of their neighbors, measured at the beginning of week $w$ (Table~\ref{tab:covariates}). As we iterate over all the weeks in the timeline, users performing link creations during several weeks will be added multiple times to any of the two groups. This is acceptable from an experimental design perspective~\cite{stuart10matching}: two versions of the same user profile at different times will have different vector of covariates, thus we will effectively consider them as two distinct user instances.

After the two groups are built, we execute the algorithm described in the previous section (\S\ref{sec:network:matching}) to obtain two statistically balanced groups. The matching algorithm yielded a pair of balanced groups with $SB<0.25$ for all covariates and an average $SB$ of 0.18. We then compare the two groups on an \textit{outcome variable} that reflects our research question. For every user instance $i$ in $G_t$ or $G_c$, we measure the quality variation of its produced content after the link creation event. This is done by computing the ratio $\Delta_b$ between the beauty of the user's photos uploaded in the week after the link creation ($b^{w+1}(i)$) and the average beauty of all its photos posted prior to the link creation event ($\bar{b}^w(i)$). When averaged over all the elements the group, the outcome variable is defined as follows:
\begin{equation}
\Delta_b(G_t) = \frac{1}{|G_t|} \cdot \sum_{i \in G_t}{\frac{b^{w+1}(i)}{\bar{b}^w(i)}} ; \textrm{(same for } G_c\textrm{)}.
\end{equation}
Figure~\ref{fig:scheme} depicts a simplified sketch of the matching experiment.

\begin{figure}[t!]
\begin{center}
\includegraphics[width=0.49\columnwidth]{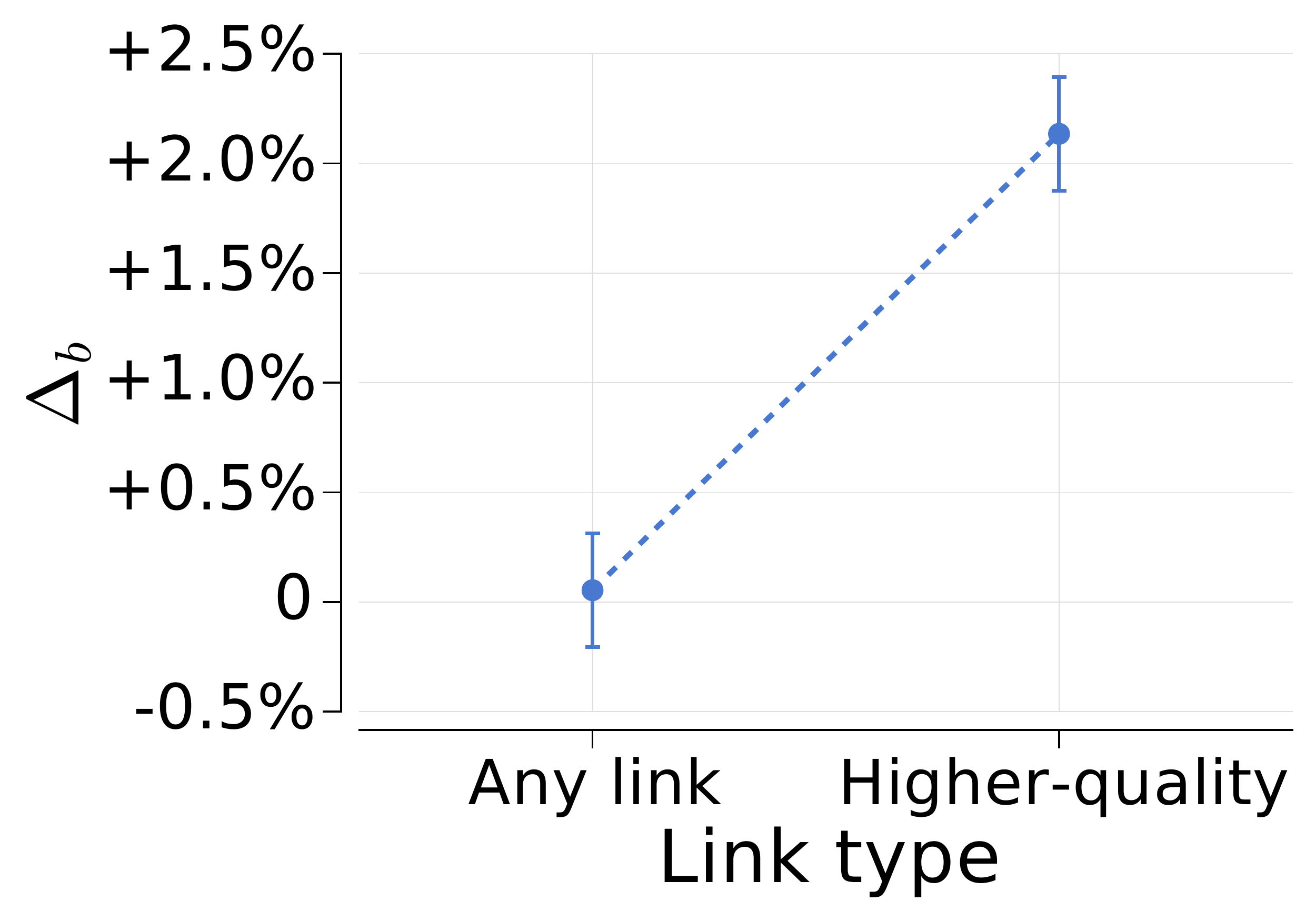}
\includegraphics[width=0.49\columnwidth]{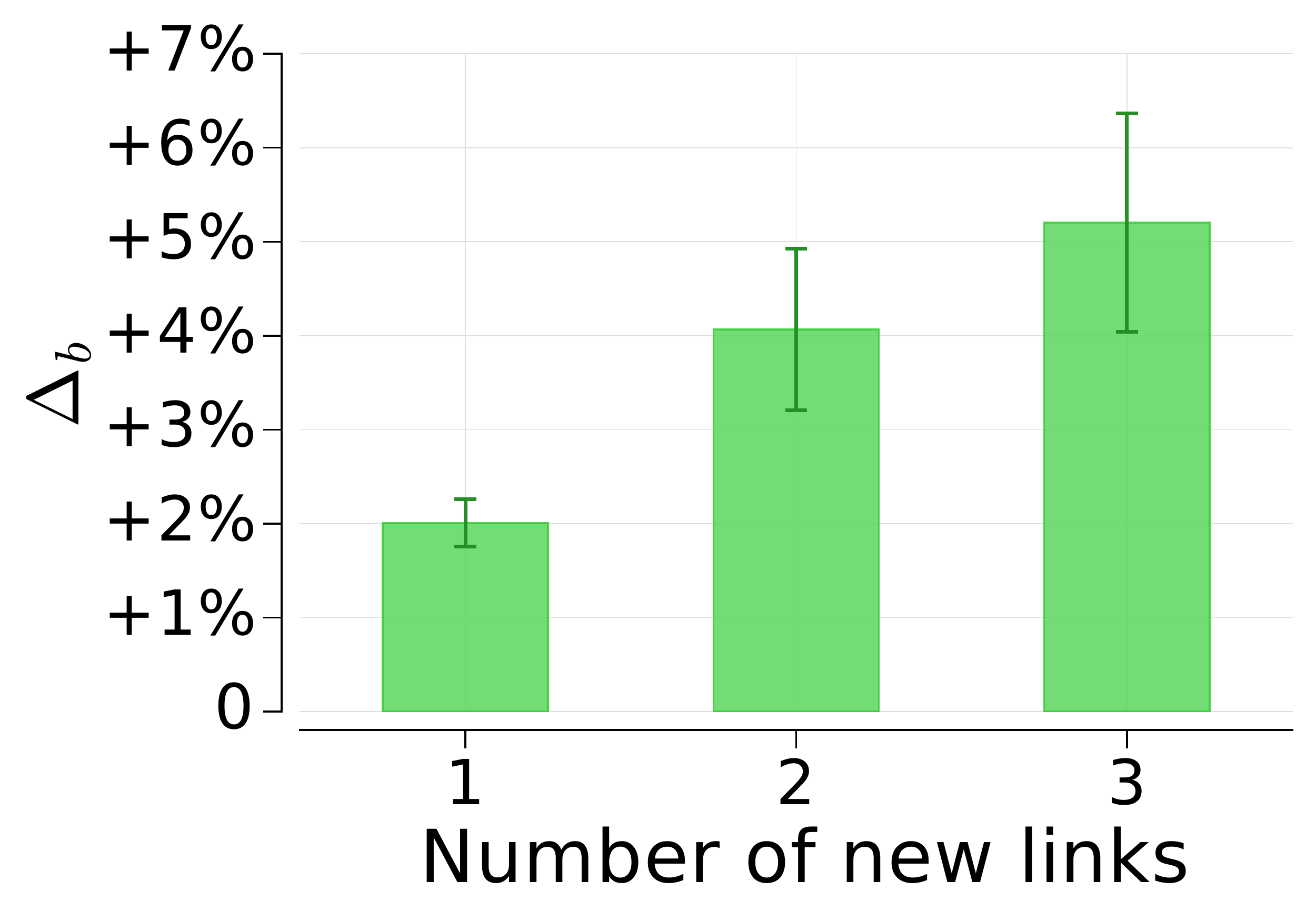}
\caption{Matching experiment. Beauty increase $\Delta_b$ after a generic link creation (control) vs. a link creation towards a user with higher beauty (treatment). Beauty increase after the creation of $n$ links towards users with higher beauty. 95\% confidence intervals are shown.}
\label{fig:matching_nlinks}
\vspace{-5mm}
\end{center}
\end{figure}

The measure of $\Delta_b$ confirms our hypothesis: the treatment group experiences an average $2\%$ increase in $\Delta_b$, whereas no significant increase is found in the control group (Figure~\ref{fig:matching_nlinks} left).

\begin{figure}[t!]
\begin{center}
\includegraphics[width=0.80\columnwidth]{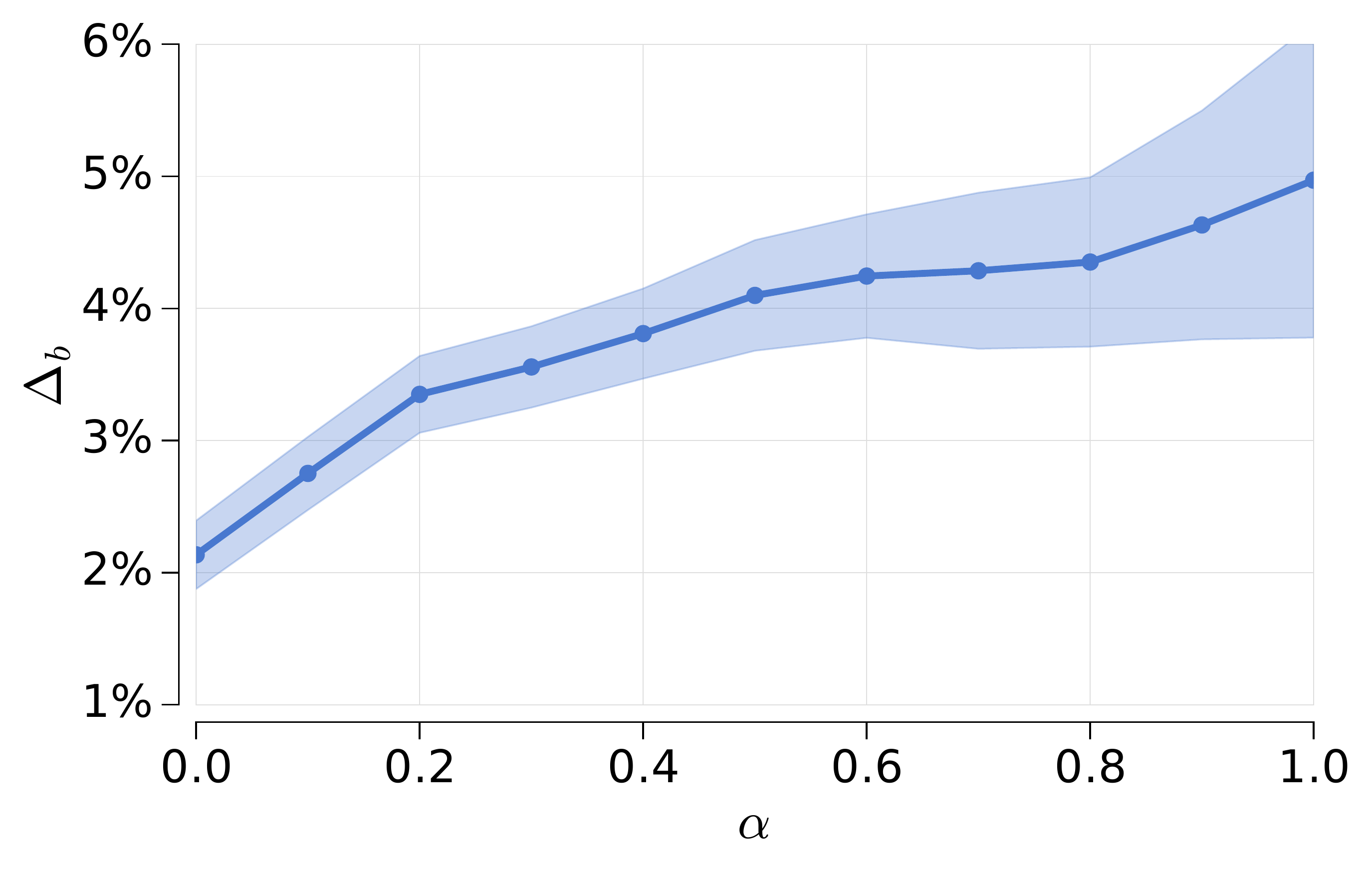}
\caption{Beauty increase of a user after creation of links towards users with quality $\alpha$ times higher. 95\% confidence interval is shown.}
\label{fig:matching_alpha}
\vspace{-5mm}
\end{center}
\end{figure}

Using the same matching setup, we run two additional experiments with new pairs of groups. First, to assess how much the influence effect is augmented by the \textit{number} of new connections, we run another matching experiment that includes in $G_t$ only users who created \textit{exactly} $n \in \{1,2,3\}$ links towards higher-quality users. We limit ourselves to $n=3$ because for larger $n$ we could not form matching pairs of treatment and control groups large enough to ensure statistical significance. We find that the influence effect accumulates with new connections, with diminishing returns (Figure~\ref{fig:matching_nlinks}, right). Last, to measure how much the beauty increase depends on the \textit{magnitude} of the difference between the user's beauty and that of its new neighbors, we restrict $G_t$ to the users whose new neighbors at week $w$ ($\Gamma^w(i)$) have an average beauty that is $\alpha$ times greater than their own:
\begin{equation}
\bar{b}^w(\Gamma^w(i)) = \frac{1}{|\Gamma^w(i)|} \cdot \sum_{j \in \Gamma^w(i)}{\bar{b}^w(j)},
\label{eq:neighbor_beauty_1}
\end{equation}
\begin{equation*}
\bar{b}^w(\Gamma^w(i))  \geq (1+\alpha) \cdot \bar{b}^w(i).
\end{equation*}
We find that, the greater the beauty differential, the greater the increase ---noticeable until $\alpha=0.5$, after which the confidence interval becomes too wide to make any assessment (Figure~\ref{fig:matching_alpha}).

In summary, we found that \textbf{A4:} \textit{users' produced quality increases as a result of new established connections with higher-quality users; the higher the number of those new contacts and the higher their quality, the stronger the effect}.

\begin{figure}[t!]
\begin{center}
\includegraphics[width=0.80\columnwidth]{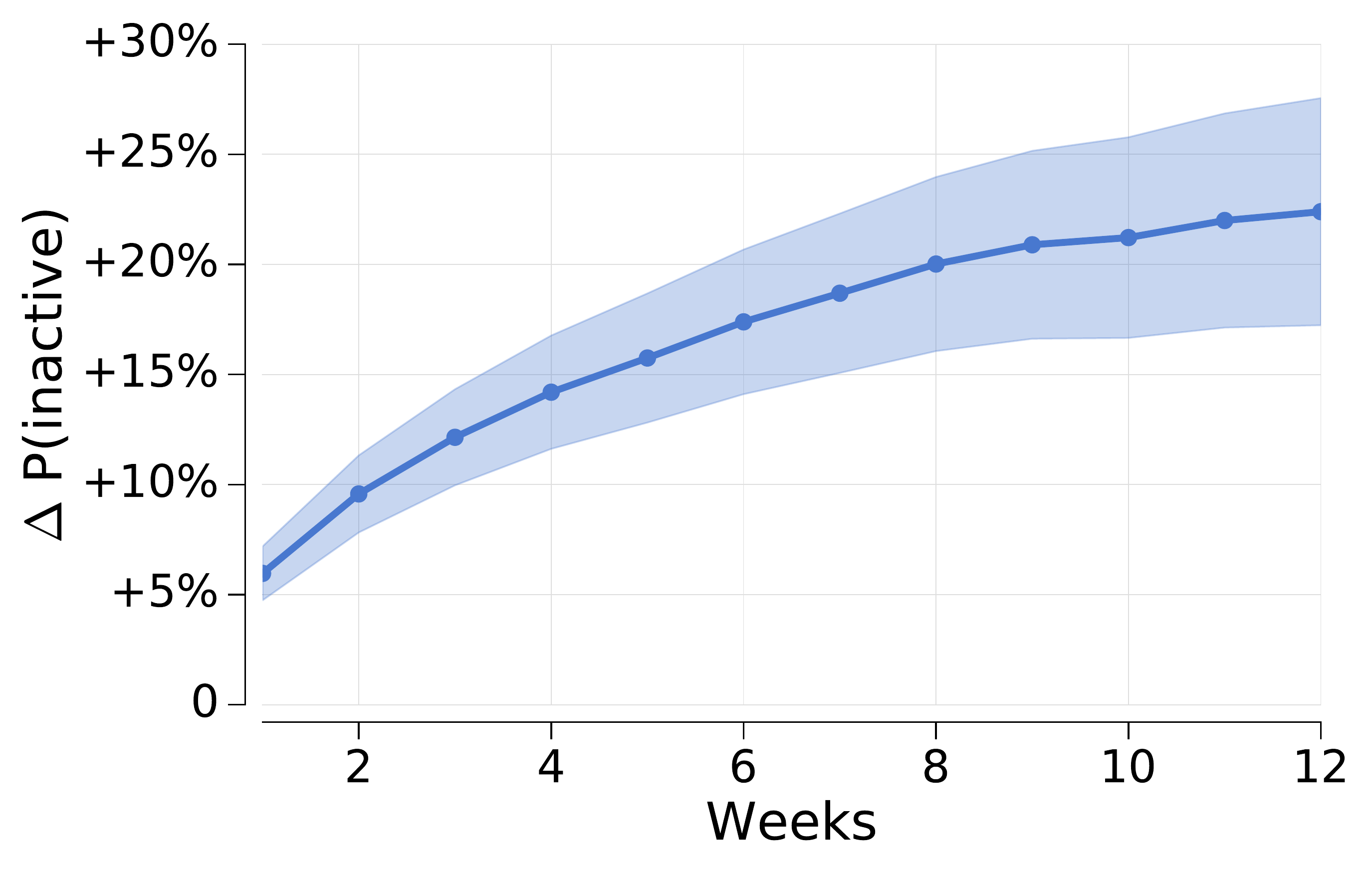}
\caption{Increase in the probability of becoming inactive for $n$ weeks for users with high beauty imbalance with their neighbors, compared to balanced users. 95\% confidence interval is shown.}
\label{fig:matching_pchurn}
\end{center}
\end{figure}

\vspace{5pt}\noindent\textbf{Beauty imbalance kills}. Finally, to answer question \textbf{Q5}, we set up an experiment to ascertain if strong quality imbalance reduces user engagement. Also for this experiment we use a weekly-quantized timeline, but this time we partition users among $G_c$ and $G_t$ based on their existing neighbor set rather than on the new connections they create. For every week-user pair $(w,i)$ we measure the average beauty of $i$'s full neighbor set at week $w$, namely $\bar{b}^w(\Gamma^w(i))$ as defined in Equation~\ref{eq:neighbor_beauty_2}. We measure how much the average neighbor beauty deviates from the user beauty
\begin{equation}
\bar{b}^w(i) + \delta \cdot \bar{b}^w(i) = \bar{b}^w(\Gamma^w(i)) 
\label{eq:neighbor_beauty_2}
\end{equation}
If the two quantities are in the same close range ($-0.1 \leq \delta \leq 0.1$), we add the user to $G_c$. Else, if the difference is substantial ---namely $30\%$ or more ($\delta \geq 0.3$)--- we add it to $G_t$. We then measure the proportion $p^{inact}$ of users in each group who remain inactive (i.e., no photo uploads) for $n$ consecutive weeks after week $w$ and compute the ratio between the values for the two sets ($p^{inact}_t/p^{inact}_c$) to measure the relative increment in treatment over control. We observe that the treatment group has higher probability of inactivity that grows from $+5\%$ to $+20\%$ in the first 12 weeks (Figure~\ref{fig:matching_pchurn}). In conclusion, we find that \textbf{A5:} \textit{people exposed to photos that deviate too much, in terms of quality, from their own contributed content, are more likely to become disengaged in the future.}

%===========================================
%===========================================
\section{Beauty-based Link Recommender} \label{sec:application}
%===========================================
%===========================================

Classic link recommendation approaches based on the graph structure (e.g., common neighbors and all its variations) tend to suggest popular and very connected users~\cite{su16effect}, thus increasing the linkage to---and consequently the level of attention on---already well-regarded individuals, keeping potential new talents away from the spotlight. However, since connectivity and user quality are largely orthogonal, algorithms that favor highly-connected users won't necessarily provide adequate visibility to high-quality content.

This point is made particularly evident if we group users according to the combination of their popularity and produced quality. We cluster Flickr users according to three variables: content quality (average beauty of the user's photos), popularity (average number of favorites per photo), and connectivity (number of followers). Given the diversity in terms of range and distributions of such variables, we first log-transform their values and then normalize them to the range $[0,1]$. Next, to identify groups of photographers with similar characteristics, we use K-means clustering over these dimensions. We vary K from 2 to 10, and select $K=4$ according to the gap statistic~\cite{tibshirani2001estimating}. The cluster centroids are reported in Table~\ref{tab:clusters}. Four classes of users emerge:

\begin{enumerate}
\setlength\itemsep{-2pt}
\item \textbf{Low Quality:} The biggest cluster contains almost half of the users. It corresponds to the long tail of ``beginner photographers'' who produce average-to-low quality content, with limited activity and low connectivity in the network.
\item \textbf{Forlorn Beauty:} The second biggest cluster gathers excellent photographers (highest average beauty value among the clusters considered) who receive very little attention from other Flickr users.
\item \textbf{Regular Users:} The regular semi-professional photographer on Flickr, sharing average-to-high quality pictures. These users are characterized by a moderate popularity within the network.
\item \textbf{Flickr Superstars:}  The smallest cluster groups together all those professional photographers (beauty level similar to the \emph{Forlorn Beauty} cluster) who are the foundation 
of the Flickr network, with many favorites and followers. Typically, these \emph{Superstars} are the ones who appear in showcase pages such as the Flickr Explore\footnote{\small \url{https://www.flickr.com/explore}}.
\end{enumerate}

\begin{table}[]
\centering
\caption{Clustering results. Photographers are divided into four groups based on their quality, popularity, and connectivity. The normalized values of those three dimensions for the four centroids are reported.}
\label{tab:clusters}
\resizebox{\columnwidth}{!}{
\begin{tabular}{l|cccc}
\multicolumn{1}{c|}{\textit{\textbf{}}} & \textbf{\%users} & \textbf{beauty} & \textbf{fav/photo} & \textbf{connects} \\ \hline
\textit{Low quality}                    & 41.2\%           & 0.17            & 0.00                     & 0.06                  \\
\textit{Forlorn beauty}                 & 28.1\%           & 0.42            & 0.01                     & 0.10                  \\              
\textit{Regular user}                   & 22.1\%          & 0.25            & 0.01                     & 0.21                  \\
\textit{Superstar}                      & 8.6\%            & 0.42            & 0.15                     & 0.35   
\end{tabular}}
\end{table}

\begin{table}[]
\centering
\caption{Average value of descriptive metrics for users in different clusters}
\label{tab:metrics}
\resizebox{\columnwidth}{!}{
\begin{tabular}{l|cccc}
\textit{\textbf{}}        & \textit{Low} & \textit{Forlorn} & \textit{Regular} & \textit{Superstar} \\ \hline
\textbf{photo count}        & 1060      & 200.4           & 1869          & 822.4             \\
\textbf{time on platform} & 104.4       & 84.68            & 187.0           & 198.3                  
\end{tabular}}
\end{table}

The clustering results confirm that the talent of a large portion of the user-base ---more that $1/4^{th}$ of the overall population--- remains largely untapped, despite its high skill level (as evidenced by the high average beauty value). This group of users is associated with a lower \emph{time on platform}, measured as the number of weeks with at least one photo upload (Table~\ref{tab:metrics}). This gives further support to the intuition that photographers who do not receive adequate recognition for their contributed value tend to churn out sooner. Furthermore, their activity in terms of number of pictures uploaded is limited (the lowest compared to other user classes), thus reducing the flow of incoming high-quality content in the platform.

Link recommender systems oblivious to quality will disproportionately recommend \textit{Superstar} users because they are very popular and well-connected. By doing so, users will be exposed to new appealing pictures because recommended contacts produce beautiful photos on average. However, this strategy has two major limitations. First, it reinforces the rich-get-richer phenomenon, depriving the users in the \textit{Forlorn beauty} class of the attention they deserve by directing it all to the small core of popular users. Last, it worsen the risk of very imbalanced connections: users who post lower-quality pictures will be mainly recommended contacts with considerably higher beauty. This is an undesirable outcome because, as we have shown earlier, accumulating many unbalanced connections increases the risk of inactivity and churn-out.

Next, building on our previous findings, we contribute to address these limitations by sketching a simple link recommendation strategy that \textit{i)} rebalances the distribution of attention to give recognition to valuable contributors otherwise forgotten, and \textit{ii)} increases the chances of a user to access new high-quality content without aggravating the quality imbalance between producers and consumers, which might cause engagement to drop in the long term.

To test this idea, we simulate a link recommendation task. We compare a classic friend-of-friend approach that recommends the contact with the highest number of common neighbors ($CN$) with an alternative, quality-oriented algorithm that recommends the user at network distance 2 with the highest average beauty score ($BB_{\pm10}$) that is within a small range from the user beauty of the recommendation recipient ($\pm$10\%), in order to avoid quality imbalance. We simulate both approaches on a random sample of 400K photographers; each of them receives only one recommendation from each approach.

\begin{figure}[t!]
\begin{center}
\includegraphics[width=0.24\columnwidth]{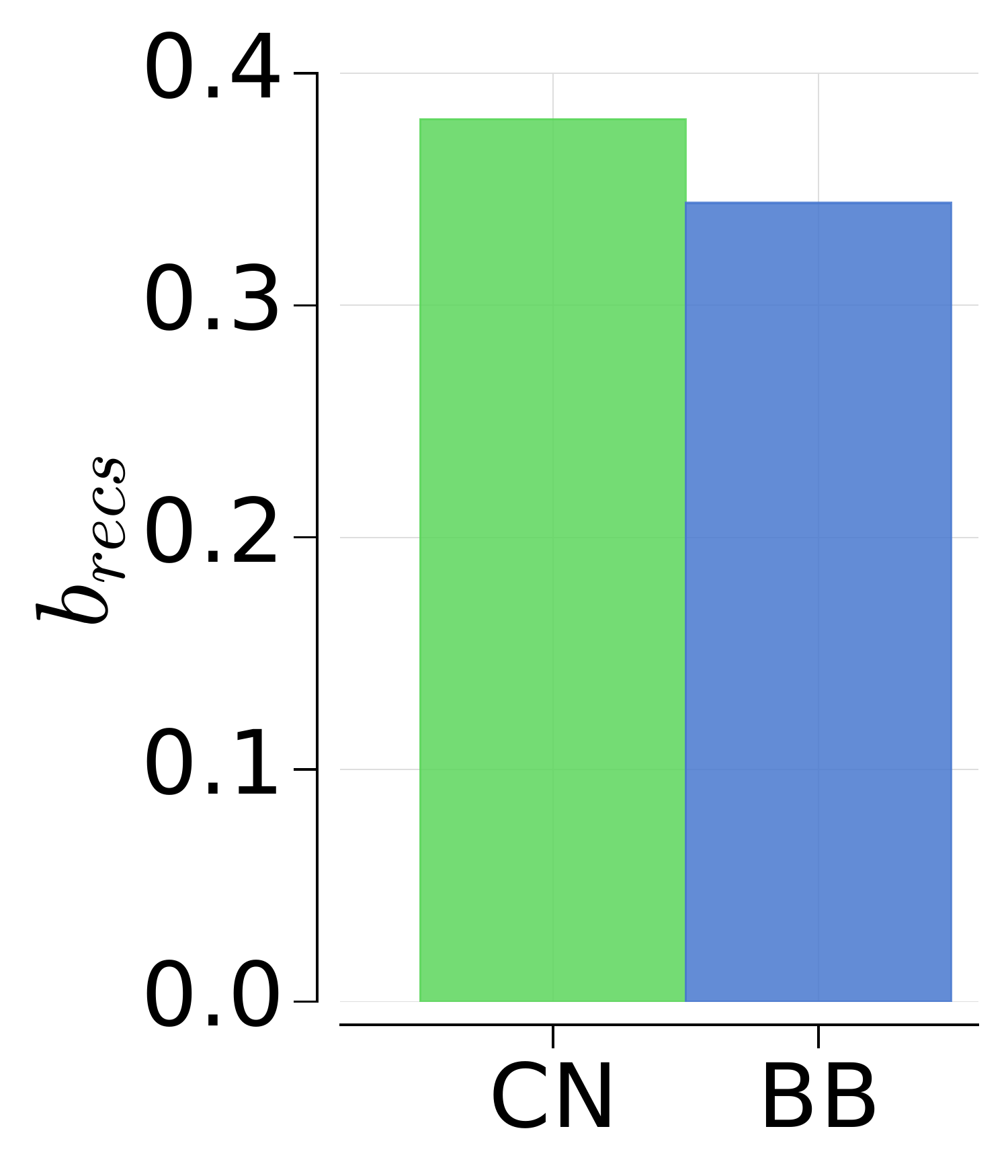}
\includegraphics[width=0.24\columnwidth]{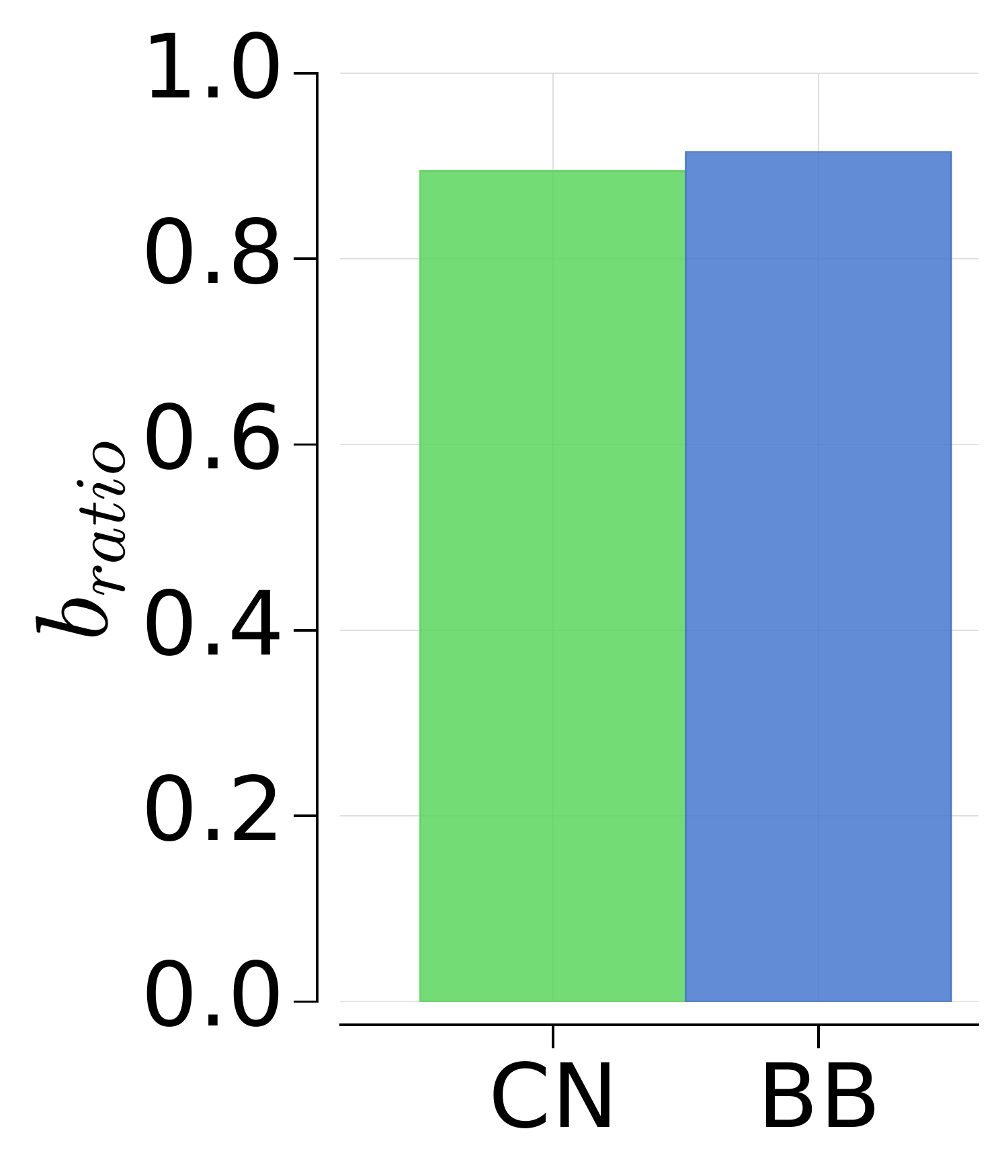}
\includegraphics[width=0.24\columnwidth]{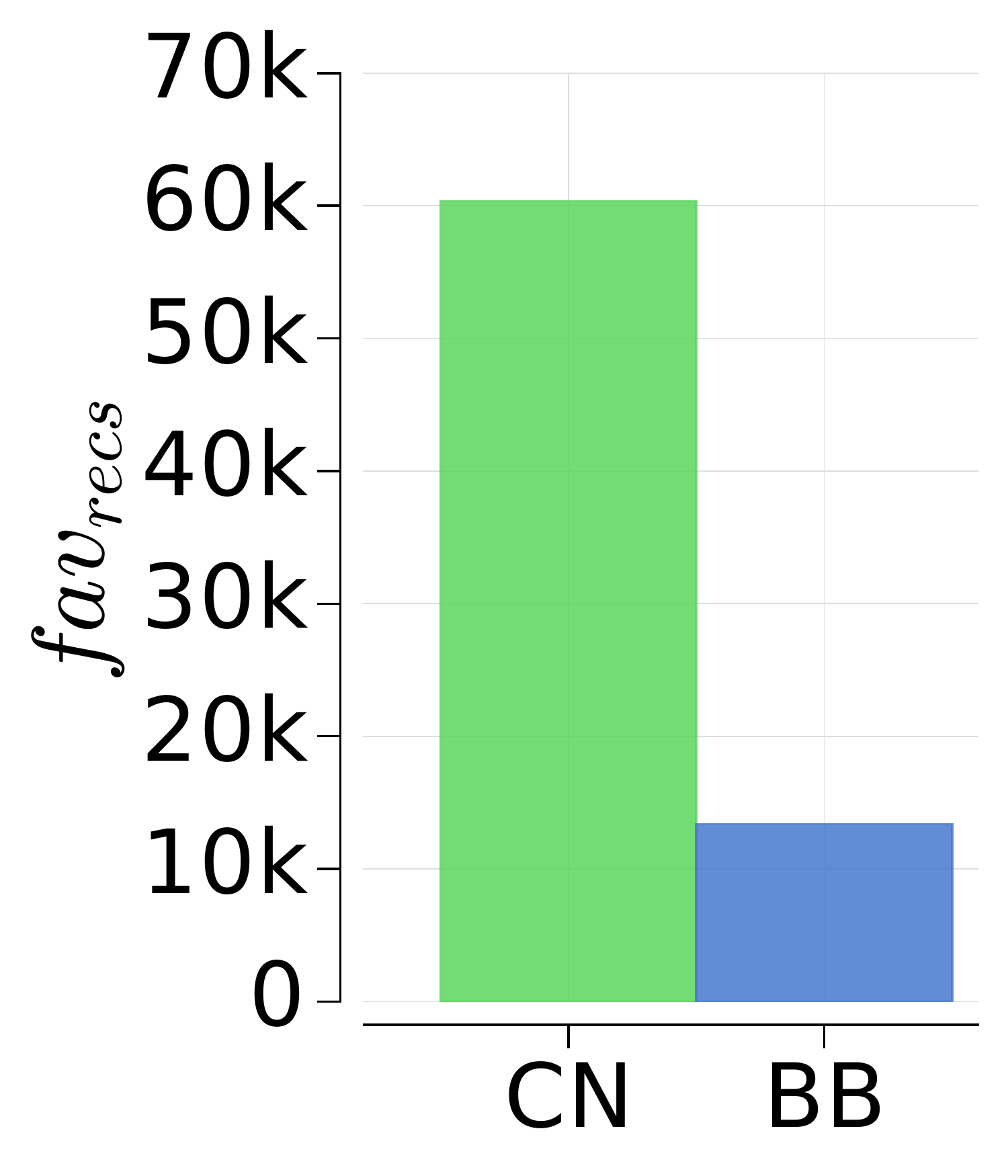}\includegraphics[width=0.24\columnwidth]{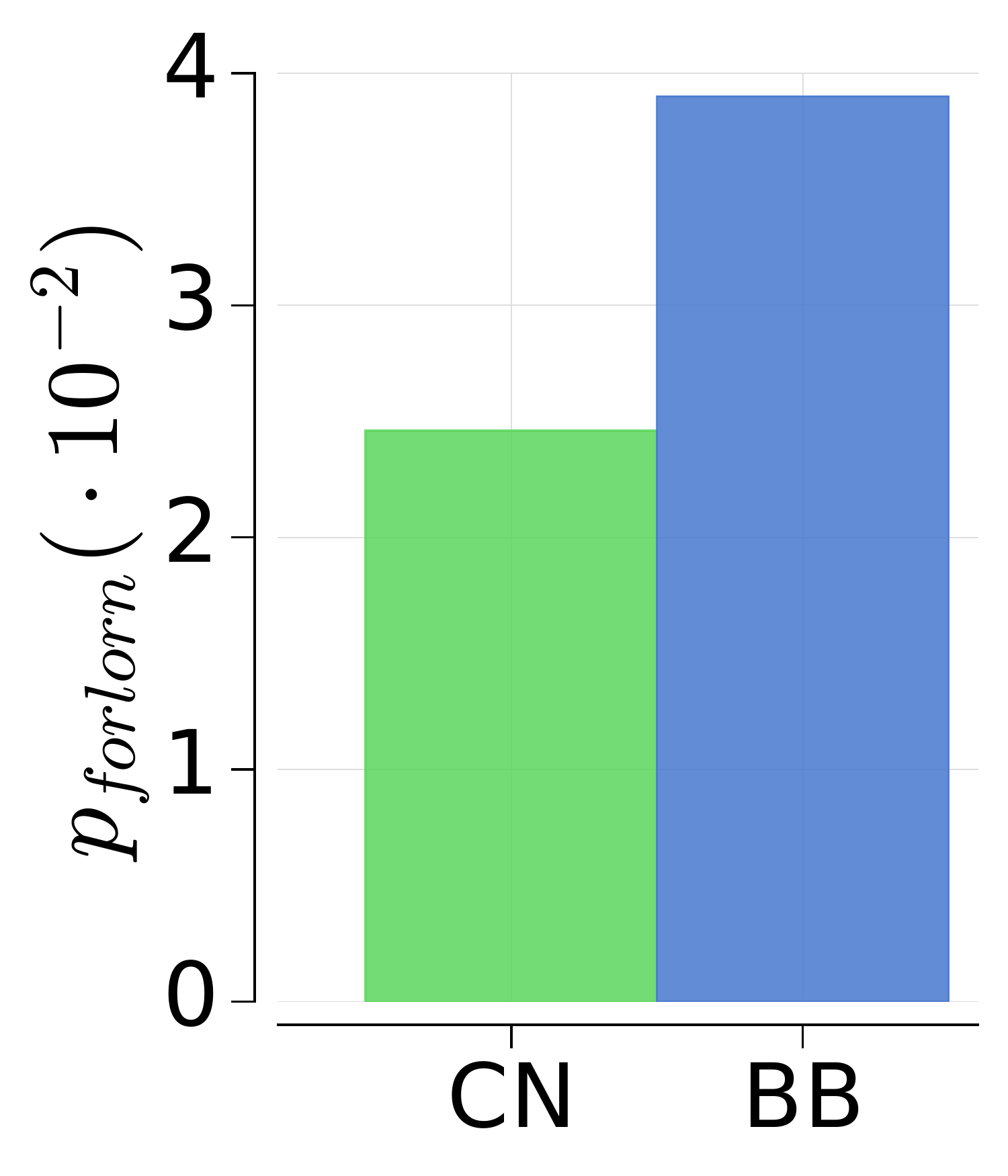}
\caption{Comparison between a friend-of-friend approach based on common neighbors (CN) and a quality-oriented algorithm that suggests users at distance 2 with the highest average beauty score ($BB_{\pm10}$) in a $\pm10\%$ quality interval.}
\label{fig:application}
\vspace{-5mm}
\end{center}
\end{figure}

Let us define $u$ as the generic user who receives the recommendation, $r$ the recommended contact, and $R$ the list of recommendations $(u,r)$. We compare the two approaches on the four indicators listed below.
\begin{itemize}
\item Average user beauty of recommended contacts \newline $b_{recs} = \frac{1}{|R|} \sum_{(u,r) \in R} \bar{b}(r)$.
\item Average ratio between the user beauty of the recommendation recipient and the user beauty of the recommended contact $b_{ratio} = \frac{1}{|R|} \sum_{(u,r) \in R} \frac{\bar{b}(u)}{\bar{b}(r)}$; a value of $b_{ratio}$ closer to 1 means a lower beauty imbalance since the two quantities in the fraction are closer.
\item Average number of favorites of recommended contacts \newline  $fav_{recs} = \frac{1}{|R|} \sum_{(u,r) \in R} fav(r)$.
\item Oortion of users in the \textit{Forlorn Beauty} cluster in the recommendation list $p_{forlorn} = \frac{\sum_{(u,r) \in R} \textbf{I} (r \in ForlornBeautySet)}{|R|}$, where $\textbf{I}(\bullet)=1$ if the condition of its argument is true, 0 otherwise.
\end{itemize}
Following the findings from $\S$6, we want to keep such imbalance low to avoid user churns on the long term. Moreover, having a higher ratio of \textit{Forlorn Beauty} users in $p_{forlorn}$ increases the exposure and potentially the future engagement of these high-quality photographers with little social attention. Figure~\ref{fig:application} shows the results of the two approaches.

The $CN$ approach selects contacts with beauty higher  than $BB_{\pm10}$, but only slightly higher, considering the strict $\pm$10\% constraint in $BB_{\pm10}$, which, by definition, will limit  the maximum level of beauty for new contacts of a given node. On the other hand, $CN$'s recommended users are 5 times more popular in terms of number of favorites. This introduces higher beauty imbalance ($+2.35\%$) and does not reach \textit{Forlorn} users as effectively: $BB_{\pm10}$ suggests 49\% more users in that class, comparatively.

Even though such a simple algorithm is far from being a production-ready solution, the simulation experiment provides initial evidence that better balance in the content consumption dynamics could be easily introduced by complementing current systems with quality-based rules.

%===========================================
%===========================================
\section{Discussion and Conclusions} \label{sec:conclusions}
%===========================================
%===========================================

Using a novel deep learning computer vision model trained on a vast image corpus from Flickr, we have conducted the first large-scale study on the relationship content quality with the social network structure.

\subsection{Implications}

Adopting popularity-driven policies to promote content and users in social networks is a fallacious way of growing healthy online communities~\cite{cha10measuring}. Nevertheless, for several years popularity has been one of the core elements of several online services including search, promoted content, and recommendations. For the first time, we have shown that it is possible to run at scale a reliable profiling of users that captures their contributed quality rather than their popularity. This can have direct practical impact not only in recommender systems, but in any application that need to retrieve, rank, or present images. Furthermore, our study about the notion of quality in combination to the network structure yields important theoretical implications in the domain of social network analysis and, more broadly, network science. We have shown that social relationships are not only homophilous, but tend also to be balanced in terms of the quality that the two endpoints produce. In line with the principles of the Social Exchange Theory, we provide empirical evidence that users who entertain strongly imbalanced social relationships in terms of the quality produced increase the risk of becoming inactive or churn out in the future. As we have shown in a simple proof of concept, next-generation link recommender systems could easily factor in the notion of quality imbalance to foster the creation of longer-lasting social ties.

\subsection{Biases}

The outcome of both the annotation task and the automated beauty scoring can be influenced by several types of biases.

We have developed the aesthetic scoring system by fine-tuning an existing neural network used for object detection. This choice is justified by computational efficiency, has been adopted in previous work, and complies with photographic theory on subject-specific aesthetic rules. Even though the image set we use to train our neural network is very large and diverse in terms of subjects, quality, and photographers, it may still contain biases that could be smoothened out by extending the training phase to multiple datasets of different nature. In future work we plan to conduct a more systematic evaluation of the biases that this approach might introduce when scoring pictures of different subjects.

The evaluation of image quality through online crowdsourcing might be affected by a number of unconscious biases originating by the personal and cultural background of the raters, the way the interface is presented, and the different subjects depicted in the photos. Although we have used a state-of-the-art framework to account for all these potential problems, a more thorough study focusing on residual biases would be desirable.
	
\subsection{Limitations and future work}

Our analysis scratches only the surface of this mostly unexplored research area. 

Our causal inference analysis groups together similar users to get a balanced matching between control and treatment sets. That is convenient to measure causal effects globally but does not directly allow for a fine-grained analysis of how meaningful user groups (e.g., newcomers vs. professional users) are impacted. The extent to which the exposure to content quality has a different impact on those user categories is an interesting extension of this work. 

The deep learning algorithm we use is very powerful but lacks explainability: in contrast with classic image aesthetic frameworks based on compositional features, it is not possible to determine why a picture has a given beauty score. Research in explainability in deep learning is still at an early stage, also in the sub-field of image aesthetics. Expanding the ability of our method to provide human-readable explanations of the beauty score is part of our planned future work.

We have described user quality with a single numeric indicator; multidimensional descriptors could add nuances to the characterization. We have studied the effect of link creation and nearest neighbors on the process of quality production; exploring a wider range of social structures and events could lead to further findings. Our experiments can determine the cause of some network dynamics (e.g., lower user engagement) but cannot provide reliable explanations about \textit{why} those changes occur; further investigation, possibly including qualitative methods, could provide more clarity on the these dynamics. Last, our experimental setting unveils causality but it is not flexible enough to reveal changes in user quality over long periods of time. Our matching strategy is effective in comparing the effect of an event (e.g., link creation) on outcome variables measured right after the event occurs, but is not designed to study long-term effects. Even though treatment and control groups are checked to be statistically equivalent over all covariates at time $t$, the likelihood that their equivalence is preserved after $t$ drops as time passes and this is why, to draw meaningful causal conclusions, it is safe to study only those outcomes (e.g., variation of user beauty) that occur right after $t$. 
As a direct consequence, it becomes hard to provide a tangible interpretation in terms of user perception of some of the small, yet significant, short-term influence effects we have found (e.g., +2\% in produced photo quality). In the future, we aim at applying more complex frameworks that can provide reliable causal inference on longer time spans.

\vspace{5pt}\noindent Despite such limitations, we hope our work contributes to a better understanding of the evolutionary dynamics of social ecosystems.

\bibliographystyle{ACM-Reference-Format}
\bibliography{bibliography}


\begin{thebibliography}{72}

%%% the Web address.

\ifx \showCODEN    \undefined \def \showCODEN     #1{\unskip}     \fi
\ifx \showDOI      \undefined \def \showDOI       #1{#1}\fi
\ifx \showISBNx    \undefined \def \showISBNx     #1{\unskip}     \fi
\ifx \showISBNxiii \undefined \def \showISBNxiii  #1{\unskip}     \fi
\ifx \showISSN     \undefined \def \showISSN      #1{\unskip}     \fi
\ifx \showLCCN     \undefined \def \showLCCN      #1{\unskip}     \fi
\ifx \shownote     \undefined \def \shownote      #1{#1}          \fi
\ifx \showarticletitle \undefined \def \showarticletitle #1{#1}   \fi
\ifx \showURL      \undefined \def \showURL       {\relax}        \fi
% invisible to TeX
\providecommand\bibfield[2]{#2}
\providecommand\bibinfo[2]{#2}
\providecommand\natexlab[1]{#1}
\providecommand\showeprint[2][]{arXiv:#2}

\bibitem[\protect\citeauthoryear{Aiello, Barrat, Schifanella, Cattuto,
  Markines, and Menczer}{Aiello et~al\mbox{.}}{2012}]%
        {aiello12tweb}
\bibfield{author}{\bibinfo{person}{Luca~Maria Aiello}, \bibinfo{person}{Alain
  Barrat}, \bibinfo{person}{Rossano Schifanella}, \bibinfo{person}{Ciro
  Cattuto}, \bibinfo{person}{Benjamin Markines}, {and} \bibinfo{person}{Filippo
  Menczer}.} \bibinfo{year}{2012}\natexlab{}.
\newblock \showarticletitle{Friendship prediction and homophily in social
  media}.
\newblock \bibinfo{journal}{{\em ACM TWEB\/}} \bibinfo{volume}{6},
  \bibinfo{number}{2}, Article \bibinfo{articleno}{9} (\bibinfo{date}{Jun}
  \bibinfo{year}{2012}).
\newblock
\showURL{%
\url{http://doi.acm.org/10.1145/2180861.2180866}}


\bibitem[\protect\citeauthoryear{Althoff, Jindal, and Leskovec}{Althoff
  et~al\mbox{.}}{2017}]%
        {althoff17online}
\bibfield{author}{\bibinfo{person}{Tim Althoff}, \bibinfo{person}{Pranav
  Jindal}, {and} \bibinfo{person}{Jure Leskovec}.}
  \bibinfo{year}{2017}\natexlab{}.
\newblock \showarticletitle{Online Actions with Offline Impact: How Online
  Social Networks Influence Online and Offline User Behavior}. In
  \bibinfo{booktitle}{{\em WSDM}}. \bibinfo{publisher}{ACM},
  \bibinfo{address}{New York, NY, USA}, \bibinfo{pages}{537--546}.
\newblock
\showISBNx{978-1-4503-4675-7}
\showDOI{%
\url{https://doi.org/10.1145/3018661.3018672}}


\bibitem[\protect\citeauthoryear{Anagnostopoulos, Kumar, and
  Mahdian}{Anagnostopoulos et~al\mbox{.}}{2008}]%
        {anagnostopoulos08influence}
\bibfield{author}{\bibinfo{person}{Aris Anagnostopoulos}, \bibinfo{person}{Ravi
  Kumar}, {and} \bibinfo{person}{Mohammad Mahdian}.}
  \bibinfo{year}{2008}\natexlab{}.
\newblock \showarticletitle{Influence and Correlation in Social Networks}. In
  \bibinfo{booktitle}{{\em KDD}}. \bibinfo{publisher}{ACM}.
\newblock
\showISBNx{978-1-60558-193-4}
\showDOI{%
\url{https://doi.org/10.1145/1401890.1401897}}


\bibitem[\protect\citeauthoryear{Aral, Muchnik, and Sundararajan}{Aral
  et~al\mbox{.}}{2009}]%
        {aral09distinguishing}
\bibfield{author}{\bibinfo{person}{Sinan Aral}, \bibinfo{person}{Lev Muchnik},
  {and} \bibinfo{person}{Arun Sundararajan}.} \bibinfo{year}{2009}\natexlab{}.
\newblock \showarticletitle{Distinguishing influence-based contagion from
  homophily-driven diffusion in dynamic networks}.
\newblock \bibinfo{journal}{{\em PNAS\/}} \bibinfo{volume}{106},
  \bibinfo{number}{51} (\bibinfo{year}{2009}).
\newblock


\bibitem[\protect\citeauthoryear{Bakshy, Rosenn, Marlow, and Adamic}{Bakshy
  et~al\mbox{.}}{2012}]%
        {bakshy12role}
\bibfield{author}{\bibinfo{person}{Eytan Bakshy}, \bibinfo{person}{Itamar
  Rosenn}, \bibinfo{person}{Cameron Marlow}, {and} \bibinfo{person}{Lada
  Adamic}.} \bibinfo{year}{2012}\natexlab{}.
\newblock \showarticletitle{The Role of Social Networks in Information
  Diffusion}. In \bibinfo{booktitle}{{\em WWW}}. \bibinfo{publisher}{ACM}.
\newblock
\showISBNx{978-1-4503-1229-5}
\showDOI{%
\url{https://doi.org/10.1145/2187836.2187907}}


\bibitem[\protect\citeauthoryear{Barrat, Barthelemy, and Vespignani}{Barrat
  et~al\mbox{.}}{2008}]%
        {barrat08dynamical}
\bibfield{author}{\bibinfo{person}{Alain Barrat}, \bibinfo{person}{Marc
  Barthelemy}, {and} \bibinfo{person}{Alessandro Vespignani}.}
  \bibinfo{year}{2008}\natexlab{}.
\newblock \bibinfo{booktitle}{{\em Dynamical processes on complex networks}}.
\newblock \bibinfo{publisher}{Cambridge University Press}.
\newblock


\bibitem[\protect\citeauthoryear{Bland and Altman}{Bland and Altman}{1997}]%
        {bland1997statistics}
\bibfield{author}{\bibinfo{person}{J~Martin Bland} {and}
  \bibinfo{person}{Douglas~G Altman}.} \bibinfo{year}{1997}\natexlab{}.
\newblock \showarticletitle{Statistics notes: Cronbach's alpha}.
\newblock \bibinfo{journal}{{\em Bmj\/}} \bibinfo{volume}{314},
  \bibinfo{number}{7080} (\bibinfo{year}{1997}), \bibinfo{pages}{572}.
\newblock


\bibitem[\protect\citeauthoryear{Blau}{Blau}{1964}]%
        {blau1964exchange}
\bibfield{author}{\bibinfo{person}{P.M. Blau}.}
  \bibinfo{year}{1964}\natexlab{}.
\newblock \bibinfo{booktitle}{{\em Exchange and Power in Social Life}}.
\newblock \bibinfo{publisher}{Wiley}.
\newblock
\showISBNx{9781412823159}
\showLCCN{64023827}
\showURL{%
\url{https://books.google.it/books?id=qhOMLscX-ZYC}}


\bibitem[\protect\citeauthoryear{Bouch, Kuchinsky, and Bhatti}{Bouch
  et~al\mbox{.}}{2000}]%
        {bouch00quality}
\bibfield{author}{\bibinfo{person}{Anna Bouch}, \bibinfo{person}{Allan
  Kuchinsky}, {and} \bibinfo{person}{Nina Bhatti}.}
  \bibinfo{year}{2000}\natexlab{}.
\newblock \showarticletitle{Quality is in the Eye of the Beholder: Meeting
  Users' Requirements for Internet Quality of Service}. In
  \bibinfo{booktitle}{{\em CHI}}. \bibinfo{publisher}{ACM}.
\newblock
\showISBNx{1-58113-216-6}
\showDOI{%
\url{https://doi.org/10.1145/332040.332447}}


\bibitem[\protect\citeauthoryear{Ceaparu, Lazar, Bessiere, Robinson, and
  Shneiderman}{Ceaparu et~al\mbox{.}}{2004}]%
        {ceaparu04determining}
\bibfield{author}{\bibinfo{person}{Irina Ceaparu}, \bibinfo{person}{Jonathan
  Lazar}, \bibinfo{person}{Katie Bessiere}, \bibinfo{person}{John Robinson},
  {and} \bibinfo{person}{Ben Shneiderman}.} \bibinfo{year}{2004}\natexlab{}.
\newblock \showarticletitle{Determining causes and severity of end-user
  frustration}.
\newblock \bibinfo{journal}{{\em International journal of human-computer
  interaction\/}} \bibinfo{volume}{17}, \bibinfo{number}{3}
  (\bibinfo{year}{2004}).
\newblock


\bibitem[\protect\citeauthoryear{Cha, Haddadi, Benevenuto, and Gummadi}{Cha
  et~al\mbox{.}}{2010}]%
        {cha10measuring}
\bibfield{author}{\bibinfo{person}{Meeyoung Cha}, \bibinfo{person}{Hamed
  Haddadi}, \bibinfo{person}{Fabr{\'\i}cio Benevenuto}, {and}
  \bibinfo{person}{Krishna~P. Gummadi}.} \bibinfo{year}{2010}\natexlab{}.
\newblock \showarticletitle{{Measuring user influence in Twitter: The million
  follower fallacy}}. In \bibinfo{booktitle}{{\em Proceedings of international
  AAAI Conference on Weblogs and Social}} {\em (\bibinfo{series}{ICWSM})}.
\newblock


\bibitem[\protect\citeauthoryear{Chu, Chen, and Chen}{Chu
  et~al\mbox{.}}{2013}]%
        {chu13size}
\bibfield{author}{\bibinfo{person}{Wei-Ta Chu}, \bibinfo{person}{Yu-Kuang
  Chen}, {and} \bibinfo{person}{Kuan-Ta Chen}.}
  \bibinfo{year}{2013}\natexlab{}.
\newblock \showarticletitle{Size Does Matter: How Image Size Affects Aesthetic
  Perception?}. In \bibinfo{booktitle}{{\em Proceedings of the 21st ACM
  International Conference on Multimedia}} {\em (\bibinfo{series}{MM '13})}.
  \bibinfo{publisher}{ACM}, \bibinfo{address}{New York, NY, USA},
  \bibinfo{pages}{53--62}.
\newblock
\showISBNx{978-1-4503-2404-5}
\showDOI{%
\url{https://doi.org/10.1145/2502081.2502102}}


\bibitem[\protect\citeauthoryear{Crandall, Cosley, Huttenlocher, Kleinberg, and
  Suri}{Crandall et~al\mbox{.}}{2008}]%
        {crandall08feedback}
\bibfield{author}{\bibinfo{person}{David Crandall}, \bibinfo{person}{Dan
  Cosley}, \bibinfo{person}{Daniel Huttenlocher}, \bibinfo{person}{Jon
  Kleinberg}, {and} \bibinfo{person}{Siddharth Suri}.}
  \bibinfo{year}{2008}\natexlab{}.
\newblock \showarticletitle{Feedback Effects Between Similarity and Social
  Influence in Online Communities}. In \bibinfo{booktitle}{{\em KDD}}.
  \bibinfo{publisher}{ACM}.
\newblock
\showISBNx{978-1-60558-193-4}
\showDOI{%
\url{https://doi.org/10.1145/1401890.1401914}}


\bibitem[\protect\citeauthoryear{Datta, Joshi, Li, and Wang}{Datta
  et~al\mbox{.}}{2006}]%
        {datta}
\bibfield{author}{\bibinfo{person}{Ritendra Datta}, \bibinfo{person}{Dhiraj
  Joshi}, \bibinfo{person}{Jia Li}, {and} \bibinfo{person}{James~Z. Wang}.}
  \bibinfo{year}{2006}\natexlab{}.
\newblock \showarticletitle{Studying Aesthetics in Photographic Images Using a
  Computational Approach}.
\newblock In \bibinfo{booktitle}{{\em ECCV}}. \bibinfo{publisher}{Springer}.
\newblock
\showURL{%
\url{http://dx.doi.org/10.1007/11744078_23}}


\bibitem[\protect\citeauthoryear{Dawid and Skene}{Dawid and Skene}{1979}]%
        {dawid1979maximum}
\bibfield{author}{\bibinfo{person}{Alexander~Philip Dawid} {and}
  \bibinfo{person}{Allan~M Skene}.} \bibinfo{year}{1979}\natexlab{}.
\newblock \showarticletitle{Maximum likelihood estimation of observer
  error-rates using the EM algorithm}.
\newblock \bibinfo{journal}{{\em Applied statistics\/}} (\bibinfo{year}{1979}),
  \bibinfo{pages}{20--28}.
\newblock


\bibitem[\protect\citeauthoryear{De~Angeli, Sutcliffe, and Hartmann}{De~Angeli
  et~al\mbox{.}}{2006}]%
        {DeAngeli:2006:IUA:1142405.1142446}
\bibfield{author}{\bibinfo{person}{Antonella De~Angeli},
  \bibinfo{person}{Alistair Sutcliffe}, {and} \bibinfo{person}{Jan Hartmann}.}
  \bibinfo{year}{2006}\natexlab{}.
\newblock \showarticletitle{Interaction, Usability and Aesthetics: What
  Influences Users' Preferences?}. In \bibinfo{booktitle}{{\em DIS}}.
  \bibinfo{publisher}{ACM}.
\newblock


\bibitem[\protect\citeauthoryear{Dewar, Li, and Davis}{Dewar
  et~al\mbox{.}}{2007}]%
        {dewar07photographic}
\bibfield{author}{\bibinfo{person}{Keith Dewar}, \bibinfo{person}{Wen~Mei Li},
  {and} \bibinfo{person}{Charles~H. Davis}.} \bibinfo{year}{2007}\natexlab{}.
\newblock \showarticletitle{Photographic Images, Culture, and Perception in
  Tourism Advertising}.
\newblock \bibinfo{journal}{{\em Journal of Travel and Tourism Marketing\/}}
  \bibinfo{volume}{22}, \bibinfo{number}{2} (\bibinfo{year}{2007}),
  \bibinfo{pages}{35--44}.
\newblock
\showDOI{%
\url{https://doi.org/10.1300/J073v22n02_03}}


\bibitem[\protect\citeauthoryear{Dhar, Ordonez, and Berg}{Dhar
  et~al\mbox{.}}{2011}]%
        {dhar11high}
\bibfield{author}{\bibinfo{person}{S. Dhar}, \bibinfo{person}{V. Ordonez},
  {and} \bibinfo{person}{T.~L. Berg}.} \bibinfo{year}{2011}\natexlab{}.
\newblock \showarticletitle{High Level Describable Attributes for Predicting
  Aesthetics and Interestingness}. In \bibinfo{booktitle}{{\em CVPR}}.
  \bibinfo{publisher}{IEEE}, 8.
\newblock
\showISBNx{978-1-4577-0394-2}
\showDOI{%
\url{https://doi.org/10.1109/CVPR.2011.5995467}}


\bibitem[\protect\citeauthoryear{Dobrian, Sekar, Awan, Stoica, Joseph, Ganjam,
  Zhan, and Zhang}{Dobrian et~al\mbox{.}}{2011}]%
        {Dobrian:2011:UIV:2018436.2018478}
\bibfield{author}{\bibinfo{person}{Florin Dobrian}, \bibinfo{person}{Vyas
  Sekar}, \bibinfo{person}{Asad Awan}, \bibinfo{person}{Ion Stoica},
  \bibinfo{person}{Dilip Joseph}, \bibinfo{person}{Aditya Ganjam},
  \bibinfo{person}{Jibin Zhan}, {and} \bibinfo{person}{Hui Zhang}.}
  \bibinfo{year}{2011}\natexlab{}.
\newblock \showarticletitle{Understanding the Impact of Video Quality on User
  Engagement}. In \bibinfo{booktitle}{{\em SIGCOMM}}. \bibinfo{publisher}{ACM}.
\newblock
\showISBNx{978-1-4503-0797-0}
\showDOI{%
\url{https://doi.org/10.1145/2018436.2018478}}


\bibitem[\protect\citeauthoryear{Feld}{Feld}{1991}]%
        {feld91your}
\bibfield{author}{\bibinfo{person}{Scott~L Feld}.}
  \bibinfo{year}{1991}\natexlab{}.
\newblock \showarticletitle{Why your friends have more friends than you do}.
\newblock \bibinfo{journal}{{\it Amer. J. Sociology}} (\bibinfo{year}{1991}).
\newblock


\bibitem[\protect\citeauthoryear{Freeman}{Freeman}{2007}]%
        {freeman2007photographer}
\bibfield{author}{\bibinfo{person}{Michael Freeman}.}
  \bibinfo{year}{2007}\natexlab{}.
\newblock \bibinfo{booktitle}{{\em The Photographer's Eye: Composition and
  Design for Better Digital Photos}}. Vol.~\bibinfo{volume}{1}.
\newblock \bibinfo{publisher}{Focal Press}.
\newblock


\bibitem[\protect\citeauthoryear{Fu, Hospedales, Xiang, Gong, and Yao}{Fu
  et~al\mbox{.}}{2014}]%
        {fleet14interestingness}
\bibfield{author}{\bibinfo{person}{Yanwei Fu}, \bibinfo{person}{TimothyM.
  Hospedales}, \bibinfo{person}{Tao Xiang}, \bibinfo{person}{Shaogang Gong},
  {and} \bibinfo{person}{Yuan Yao}.} \bibinfo{year}{2014}\natexlab{}.
\newblock \showarticletitle{Interestingness Prediction by Robust Learning to
  Rank}.
\newblock In \bibinfo{booktitle}{{\em Computer Vision -- ECCV 2014}},
  \bibfield{editor}{\bibinfo{person}{David Fleet}, \bibinfo{person}{Tomas
  Pajdla}, \bibinfo{person}{Bernt Schiele}, {and} \bibinfo{person}{Tinne
  Tuytelaars}} (Eds.). \bibinfo{series}{Lecture Notes in Computer Science},
  Vol.~\bibinfo{volume}{8690}. \bibinfo{publisher}{Springer International
  Publishing}, \bibinfo{pages}{488--503}.
\newblock
\showISBNx{978-3-319-10604-5}
\showDOI{%
\url{https://doi.org/10.1007/978-3-319-10605-2_32}}


\bibitem[\protect\citeauthoryear{Girshick, Donahue, Darrell, and
  Malik}{Girshick et~al\mbox{.}}{2014}]%
        {girshick2014rich}
\bibfield{author}{\bibinfo{person}{Ross Girshick}, \bibinfo{person}{Jeff
  Donahue}, \bibinfo{person}{Trevor Darrell}, {and} \bibinfo{person}{Jitendra
  Malik}.} \bibinfo{year}{2014}\natexlab{}.
\newblock \showarticletitle{Rich feature hierarchies for accurate object
  detection and semantic segmentation}. In \bibinfo{booktitle}{{\em CVPR}}.
\newblock


\bibitem[\protect\citeauthoryear{Gulliver and Ghinea}{Gulliver and
  Ghinea}{2006}]%
        {gulliver06defining}
\bibfield{author}{\bibinfo{person}{Stephen~R Gulliver} {and}
  \bibinfo{person}{Gheorghita Ghinea}.} \bibinfo{year}{2006}\natexlab{}.
\newblock \showarticletitle{Defining user perception of distributed multimedia
  quality}.
\newblock \bibinfo{journal}{{\em ACM TOMM\/}} \bibinfo{volume}{2},
  \bibinfo{number}{4} (\bibinfo{year}{2006}).
\newblock


\bibitem[\protect\citeauthoryear{Hagen and Jones}{Hagen and Jones}{1978}]%
        {hagen78perception}
\bibfield{author}{\bibinfo{person}{Margaret~A. Hagen} {and}
  \bibinfo{person}{Rebecca~K. Jones}.} \bibinfo{year}{1978}\natexlab{}.
\newblock \showarticletitle{Cultural Effects on Pictorial Perception: How Many
  Words Is One Picture Really Worth?}
\newblock In \bibinfo{booktitle}{{\em Perception and Experience}},
  \bibfield{editor}{\bibinfo{person}{RichardD. Walk} {and} \bibinfo{person}{Jr.
  Pick, HerbertL.}} (Eds.). \bibinfo{series}{Perception and Perceptual
  Development}, Vol.~\bibinfo{volume}{1}. \bibinfo{publisher}{Springer US},
  \bibinfo{pages}{171--212}.
\newblock
\showISBNx{978-1-4684-2621-2}
\showDOI{%
\url{https://doi.org/10.1007/978-1-4684-2619-9_6}}


\bibitem[\protect\citeauthoryear{Halvey and Keane}{Halvey and Keane}{2007}]%
        {halvey07exploring}
\bibfield{author}{\bibinfo{person}{Martin~J. Halvey} {and}
  \bibinfo{person}{Mark~T. Keane}.} \bibinfo{year}{2007}\natexlab{}.
\newblock \showarticletitle{Exploring Social Dynamics in Online Media Sharing}.
  In \bibinfo{booktitle}{{\em WWW}}. \bibinfo{publisher}{ACM}.
\newblock
\showISBNx{978-1-59593-654-7}
\showDOI{%
\url{https://doi.org/10.1145/1242572.1242804}}


\bibitem[\protect\citeauthoryear{Hodas, Kooti, and Lerman}{Hodas
  et~al\mbox{.}}{2013}]%
        {hodas13icwsm}
\bibfield{author}{\bibinfo{person}{Nathan~O. Hodas}, \bibinfo{person}{Farshad
  Kooti}, {and} \bibinfo{person}{Kristina Lerman}.}
  \bibinfo{year}{2013}\natexlab{}.
\newblock \showarticletitle{Friendship Paradox Redux: Your Friends Are More
  Interesting Than You}. In \bibinfo{booktitle}{{\em ICWSM}}.
  \bibinfo{publisher}{AAAI}.
\newblock


\bibitem[\protect\citeauthoryear{Hurter}{Hurter}{2007}]%
        {hurter2007portrait}
\bibfield{author}{\bibinfo{person}{Bill Hurter}.}
  \bibinfo{year}{2007}\natexlab{}.
\newblock \bibinfo{booktitle}{{\em Portrait Photographer's Handbook}}.
\newblock \bibinfo{publisher}{Amherst Media, Inc}.
\newblock


\bibitem[\protect\citeauthoryear{Isola, Xiao, Torralba, and Oliva}{Isola
  et~al\mbox{.}}{2011}]%
        {isola2011}
\bibfield{author}{\bibinfo{person}{Phillip Isola}, \bibinfo{person}{Jianxiong
  Xiao}, \bibinfo{person}{Antonio Torralba}, {and} \bibinfo{person}{Aude
  Oliva}.} \bibinfo{year}{2011}\natexlab{}.
\newblock \showarticletitle{What makes an image memorable?}. In
  \bibinfo{booktitle}{{\em CVPR}}. \bibinfo{publisher}{IEEE}.
\newblock


\bibitem[\protect\citeauthoryear{Jiang, Wang, Feng, Xue, Zheng, and Yang}{Jiang
  et~al\mbox{.}}{2013}]%
        {jiang2013understanding}
\bibfield{author}{\bibinfo{person}{Yu-Gang Jiang}, \bibinfo{person}{Yanran
  Wang}, \bibinfo{person}{Rui Feng}, \bibinfo{person}{Xiangyang Xue},
  \bibinfo{person}{Yingbin Zheng}, {and} \bibinfo{person}{Hanfang Yang}.}
  \bibinfo{year}{2013}\natexlab{}.
\newblock \showarticletitle{Understanding and Predicting Interestingness of
  Videos.}. In \bibinfo{booktitle}{{\em AAAI}}.
\newblock


\bibitem[\protect\citeauthoryear{Jin, Chi, Peng, Tian, Ye, and Li}{Jin
  et~al\mbox{.}}{2016}]%
        {jin16deep}
\bibfield{author}{\bibinfo{person}{Xin Jin}, \bibinfo{person}{Jingying Chi},
  \bibinfo{person}{Siwei Peng}, \bibinfo{person}{Yulu Tian},
  \bibinfo{person}{Chaochen Ye}, {and} \bibinfo{person}{Xiaodong Li}.}
  \bibinfo{year}{2016}\natexlab{}.
\newblock \showarticletitle{Deep image aesthetics classification using
  inception modules and fine-tuning connected layer}. In
  \bibinfo{booktitle}{{\em Wireless Communications \& Signal Processing (WCSP),
  2016 8th International Conference on}}. IEEE, \bibinfo{pages}{1--6}.
\newblock


\bibitem[\protect\citeauthoryear{Ke, Tang, and Jing}{Ke et~al\mbox{.}}{2006}]%
        {ke2006design}
\bibfield{author}{\bibinfo{person}{Yan Ke}, \bibinfo{person}{Xiaoou Tang},
  {and} \bibinfo{person}{Feng Jing}.} \bibinfo{year}{2006}\natexlab{}.
\newblock \showarticletitle{The design of high-level features for photo quality
  assessment}. In \bibinfo{booktitle}{{\em CVPR}}. IEEE.
\newblock


\bibitem[\protect\citeauthoryear{Kong, Shen, Lin, Mech, and Fowlkes}{Kong
  et~al\mbox{.}}{2016}]%
        {kong16photo}
\bibfield{author}{\bibinfo{person}{Shu Kong}, \bibinfo{person}{Xiaohui Shen},
  \bibinfo{person}{Zhe Lin}, \bibinfo{person}{Radomir Mech}, {and}
  \bibinfo{person}{Charless Fowlkes}.} \bibinfo{year}{2016}\natexlab{}.
\newblock \showarticletitle{Photo aesthetics ranking network with attributes
  and content adaptation}. In \bibinfo{booktitle}{{\em European Conference on
  Computer Vision}}. Springer, \bibinfo{pages}{662--679}.
\newblock


\bibitem[\protect\citeauthoryear{Krizhevsky, Sutskever, and Hinton}{Krizhevsky
  et~al\mbox{.}}{2012}]%
        {krizhevsky2012imagenet}
\bibfield{author}{\bibinfo{person}{Alex Krizhevsky}, \bibinfo{person}{Ilya
  Sutskever}, {and} \bibinfo{person}{Geoffrey~E Hinton}.}
  \bibinfo{year}{2012}\natexlab{}.
\newblock \showarticletitle{Imagenet classification with deep convolutional
  neural networks}. In \bibinfo{booktitle}{{\em NIPS}}.
\newblock


\bibitem[\protect\citeauthoryear{Lavie and Tractinsky}{Lavie and
  Tractinsky}{2004}]%
        {Lavie:2004:ADP:998271.998272}
\bibfield{author}{\bibinfo{person}{Talia Lavie} {and} \bibinfo{person}{Noam
  Tractinsky}.} \bibinfo{year}{2004}\natexlab{}.
\newblock \showarticletitle{Assessing Dimensions of Perceived Visual Aesthetics
  of Web Sites}.
\newblock \bibinfo{journal}{{\em International Journal of Human-Computer
  Studie\/}} \bibinfo{volume}{60}, \bibinfo{number}{3} (\bibinfo{year}{2004}).
\newblock


\bibitem[\protect\citeauthoryear{Lerman, Yan, and Wu}{Lerman
  et~al\mbox{.}}{2016}]%
        {lerman16majority}
\bibfield{author}{\bibinfo{person}{Kristina Lerman}, \bibinfo{person}{Xiaoran
  Yan}, {and} \bibinfo{person}{Xin-Zeng Wu}.} \bibinfo{year}{2016}\natexlab{}.
\newblock \showarticletitle{The majority illusion in social networks}.
\newblock \bibinfo{journal}{{\em PloS one\/}} \bibinfo{volume}{11},
  \bibinfo{number}{2} (\bibinfo{year}{2016}).
\newblock


\bibitem[\protect\citeauthoryear{Lu, Lin, Jin, Yang, and Wang}{Lu
  et~al\mbox{.}}{2014}]%
        {luo2014rapid}
\bibfield{author}{\bibinfo{person}{Xin Lu}, \bibinfo{person}{Zhe Lin},
  \bibinfo{person}{Hailin Jin}, \bibinfo{person}{Jianchao Yang}, {and}
  \bibinfo{person}{James~Z. Wang}.} \bibinfo{year}{2014}\natexlab{}.
\newblock \showarticletitle{RAPID: Rating Pictorial Aesthetics Using Deep
  Learning}. In \bibinfo{booktitle}{{\em Multimedia}}.
  \bibinfo{publisher}{ACM}, 10.
\newblock
\showISBNx{978-1-4503-3063-3}
\showDOI{%
\url{https://doi.org/10.1145/2647868.2654927}}


\bibitem[\protect\citeauthoryear{Luo, Wang, and Tang}{Luo
  et~al\mbox{.}}{2011}]%
        {luo11content}
\bibfield{author}{\bibinfo{person}{Wei Luo}, \bibinfo{person}{Xiaogang Wang},
  {and} \bibinfo{person}{Xiaoou Tang}.} \bibinfo{year}{2011}\natexlab{}.
\newblock \showarticletitle{Content-based photo quality assessment}. In
  \bibinfo{booktitle}{{\em ICCV}}. IEEE, \bibinfo{pages}{2206--2213}.
\newblock


\bibitem[\protect\citeauthoryear{Luo and Tang}{Luo and Tang}{2008}]%
        {luo08photo}
\bibfield{author}{\bibinfo{person}{Yiwen Luo} {and} \bibinfo{person}{Xiaoou
  Tang}.} \bibinfo{year}{2008}\natexlab{}.
\newblock \showarticletitle{Photo and Video Quality Evaluation: Focusing on the
  Subject}.
\newblock In \bibinfo{booktitle}{{\em ECCV}}. \bibinfo{publisher}{Springer}.
\newblock
\showISBNx{978-3-540-88689-1}
\showDOI{%
\url{https://doi.org/10.1007/978-3-540-88690-7_29}}


\bibitem[\protect\citeauthoryear{Machajdik and Hanbury}{Machajdik and
  Hanbury}{2010}]%
        {emotions}
\bibfield{author}{\bibinfo{person}{Jana Machajdik} {and} \bibinfo{person}{Allan
  Hanbury}.} \bibinfo{year}{2010}\natexlab{}.
\newblock \showarticletitle{Affective image classification using features
  inspired by psychology and art theory}. In \bibinfo{booktitle}{{\em
  Multimedia}}. ACM.
\newblock


\bibitem[\protect\citeauthoryear{Mai, Jin, and Liu}{Mai et~al\mbox{.}}{2016}]%
        {mai16composition}
\bibfield{author}{\bibinfo{person}{Long Mai}, \bibinfo{person}{Hailin Jin},
  {and} \bibinfo{person}{Feng Liu}.} \bibinfo{year}{2016}\natexlab{}.
\newblock \showarticletitle{Composition-preserving deep photo aesthetics
  assessment}. In \bibinfo{booktitle}{{\em CVPR, Proceedings of the IEEE
  Conference on Computer Vision and Pattern Recognition}}.
  \bibinfo{pages}{497--506}.
\newblock


\bibitem[\protect\citeauthoryear{Marchesotti, Perronnin, Larlus, and
  Csurka}{Marchesotti et~al\mbox{.}}{2011}]%
        {marchesotti2011assessing}
\bibfield{author}{\bibinfo{person}{Luca Marchesotti}, \bibinfo{person}{Florent
  Perronnin}, \bibinfo{person}{Diane Larlus}, {and} \bibinfo{person}{Gabriela
  Csurka}.} \bibinfo{year}{2011}\natexlab{}.
\newblock \showarticletitle{Assessing the aesthetic quality of photographs
  using generic image descriptors}. In \bibinfo{booktitle}{{\em ICCV}}. IEEE.
\newblock


\bibitem[\protect\citeauthoryear{Mason and Suri}{Mason and Suri}{2012}]%
        {mason14conducting}
\bibfield{author}{\bibinfo{person}{Winter Mason} {and}
  \bibinfo{person}{Siddharth Suri}.} \bibinfo{year}{2012}\natexlab{}.
\newblock \showarticletitle{Conducting behavioral research on Amazon's
  Mechanical Turk}.
\newblock \bibinfo{journal}{{\em Behavior Research Methods\/}}
  \bibinfo{volume}{44}, \bibinfo{number}{1} (\bibinfo{year}{2012}),
  \bibinfo{pages}{1--23}.
\newblock
\showDOI{%
\url{https://doi.org/10.3758/s13428-011-0124-6}}


\bibitem[\protect\citeauthoryear{Mislove, Marcon, Gummadi, Druschel, and
  Bhattacharjee}{Mislove et~al\mbox{.}}{2007}]%
        {mislove07measurement}
\bibfield{author}{\bibinfo{person}{Alan Mislove}, \bibinfo{person}{Massimiliano
  Marcon}, \bibinfo{person}{Krishna~P. Gummadi}, \bibinfo{person}{Peter
  Druschel}, {and} \bibinfo{person}{Bobby Bhattacharjee}.}
  \bibinfo{year}{2007}\natexlab{}.
\newblock \showarticletitle{Measurement and Analysis of Online Social
  Networks}. In \bibinfo{booktitle}{{\em IMC}}. \bibinfo{publisher}{ACM}, 14.
\newblock
\showISBNx{978-1-59593-908-1}
\showDOI{%
\url{https://doi.org/10.1145/1298306.1298311}}


\bibitem[\protect\citeauthoryear{Miyamoto, Nisbett, and Masuda}{Miyamoto
  et~al\mbox{.}}{2006}]%
        {miyamoto06culture}
\bibfield{author}{\bibinfo{person}{Yuri Miyamoto}, \bibinfo{person}{Richard~E.
  Nisbett}, {and} \bibinfo{person}{Takahiko Masuda}.}
  \bibinfo{year}{2006}\natexlab{}.
\newblock \showarticletitle{Culture and the Physical Environment: Holistic
  Versus Analytic Perceptual Affordances}.
\newblock \bibinfo{journal}{{\em Psychological Science\/}}
  \bibinfo{volume}{17}, \bibinfo{number}{2} (\bibinfo{year}{2006}),
  \bibinfo{pages}{113--119}.
\newblock
\showDOI{%
\url{https://doi.org/10.1111/j.1467-9280.2006.01673.x}}
\showeprint{http://pss.sagepub.com/content/17/2/113.full.pdf+html}


\bibitem[\protect\citeauthoryear{Murray, Marchesotti, and Perronnin}{Murray
  et~al\mbox{.}}{2012}]%
        {murray2012ava}
\bibfield{author}{\bibinfo{person}{Naila Murray}, \bibinfo{person}{Luca
  Marchesotti}, {and} \bibinfo{person}{Florent Perronnin}.}
  \bibinfo{year}{2012}\natexlab{}.
\newblock \showarticletitle{AVA: A large-scale database for aesthetic visual
  analysis}. In \bibinfo{booktitle}{{\em CVPR}}. IEEE.
\newblock


\bibitem[\protect\citeauthoryear{Nishiyama, Okabe, Sato, and Sato}{Nishiyama
  et~al\mbox{.}}{2011}]%
        {nishiyama2011aesthetic}
\bibfield{author}{\bibinfo{person}{Masashi Nishiyama},
  \bibinfo{person}{Takahiro Okabe}, \bibinfo{person}{Imari Sato}, {and}
  \bibinfo{person}{Yoichi Sato}.} \bibinfo{year}{2011}\natexlab{}.
\newblock \showarticletitle{Aesthetic quality classification of photographs
  based on color harmony}. In \bibinfo{booktitle}{{\em CVPR}}. IEEE.
\newblock


\bibitem[\protect\citeauthoryear{Obrador, Anguera, de~Oliveira, and
  Oliver}{Obrador et~al\mbox{.}}{2009}]%
        {obrador09role}
\bibfield{author}{\bibinfo{person}{Pere Obrador}, \bibinfo{person}{Xavier
  Anguera}, \bibinfo{person}{Rodrigo de Oliveira}, {and} \bibinfo{person}{Nuria
  Oliver}.} \bibinfo{year}{2009}\natexlab{}.
\newblock \showarticletitle{The Role of Tags and Image Aesthetics in Social
  Image Search}. In \bibinfo{booktitle}{{\em WSM}}. \bibinfo{publisher}{ACM},
  8.
\newblock
\showISBNx{978-1-60558-759-2}
\showDOI{%
\url{https://doi.org/10.1145/1631144.1631158}}


\bibitem[\protect\citeauthoryear{Obrador, Saad, Suryanarayan, and
  Oliver}{Obrador et~al\mbox{.}}{2012}]%
        {obrador2012towards}
\bibfield{author}{\bibinfo{person}{Pere Obrador}, \bibinfo{person}{Michele~A
  Saad}, \bibinfo{person}{Poonam Suryanarayan}, {and} \bibinfo{person}{Nuria
  Oliver}.} \bibinfo{year}{2012}\natexlab{}.
\newblock \bibinfo{booktitle}{{\em Towards category-based aesthetic models of
  photographs}}.
\newblock \bibinfo{publisher}{Springer}.
\newblock


\bibitem[\protect\citeauthoryear{Olteanu, Varol, and Kiciman}{Olteanu
  et~al\mbox{.}}{2017}]%
        {olteanu17distilling}
\bibfield{author}{\bibinfo{person}{Alexandra Olteanu}, \bibinfo{person}{Onur
  Varol}, {and} \bibinfo{person}{Emre Kiciman}.}
  \bibinfo{year}{2017}\natexlab{}.
\newblock \showarticletitle{Distilling the Outcomes of Personal Experiences: A
  Propensity-scored Analysis of Social Media}. In \bibinfo{booktitle}{{\em
  CSCW}}. \bibinfo{publisher}{ACM}, \bibinfo{address}{New York, NY, USA},
  \bibinfo{pages}{370--386}.
\newblock
\showISBNx{978-1-4503-4335-0}
\showDOI{%
\url{https://doi.org/10.1145/2998181.2998353}}


\bibitem[\protect\citeauthoryear{Redi, Ho{\ss}feld, Korshunov, Mazza, Povoa,
  and Keimel}{Redi et~al\mbox{.}}{2014a}]%
        {redi2013crowdsourcing}
\bibfield{author}{\bibinfo{person}{Judith~Alice Redi}, \bibinfo{person}{Tobias
  Ho{\ss}feld}, \bibinfo{person}{Pavel Korshunov}, \bibinfo{person}{Filippo
  Mazza}, \bibinfo{person}{Isabel Povoa}, {and} \bibinfo{person}{Christian
  Keimel}.} \bibinfo{year}{2014}\natexlab{a}.
\newblock \showarticletitle{Crowdsourcing-based multimedia subjective
  evaluations: a case study on image recognizability and aesthetic appeal}. In
  \bibinfo{booktitle}{{\em Sixth International Workshop on Quality of
  Multimedia Experience (QoMEX 2014)}}. \bibinfo{pages}{29--34}.
\newblock


\bibitem[\protect\citeauthoryear{Redi and Merialdo}{Redi and Merialdo}{2012}]%
        {redi12where}
\bibfield{author}{\bibinfo{person}{Miriam Redi} {and} \bibinfo{person}{Bernard
  Merialdo}.} \bibinfo{year}{2012}\natexlab{}.
\newblock \showarticletitle{Where is the Beauty?: Retrieving Appealing
  VideoScenes by Learning Flickr-based Graded Judgments}. In
  \bibinfo{booktitle}{{\em Multimedia}}. \bibinfo{publisher}{ACM}.
\newblock
\showISBNx{978-1-4503-1089-5}
\showDOI{%
\url{https://doi.org/10.1145/2393347.2396486}}


\bibitem[\protect\citeauthoryear{Redi, O'Hare, Schifanella, Trevisiol, and
  Jaimes}{Redi et~al\mbox{.}}{2014b}]%
        {redi6}
\bibfield{author}{\bibinfo{person}{Miriam Redi}, \bibinfo{person}{Neil O'Hare},
  \bibinfo{person}{Rossano Schifanella}, \bibinfo{person}{Michele Trevisiol},
  {and} \bibinfo{person}{Alejandro Jaimes}.} \bibinfo{year}{2014}\natexlab{b}.
\newblock \showarticletitle{6 seconds of sound and vision: Creativity in
  micro-videos}. In \bibinfo{booktitle}{{\em CVPR}}. IEEE.
\newblock


\bibitem[\protect\citeauthoryear{Redi, Rasiwasia, Aggarwal, and Jaimes}{Redi
  et~al\mbox{.}}{2015}]%
        {redi15thebeauty}
\bibfield{author}{\bibinfo{person}{Miriam Redi}, \bibinfo{person}{Nikhil
  Rasiwasia}, \bibinfo{person}{Gaurav Aggarwal}, {and}
  \bibinfo{person}{Alejandro Jaimes}.} \bibinfo{year}{2015}\natexlab{}.
\newblock \showarticletitle{The Beauty of Capturing Faces: Rating the Quality
  of Digital Portraits}. In \bibinfo{booktitle}{{\em IEEE International
  Conference on Automatic Face and Gesture Recognition 2015}}.
  \bibinfo{publisher}{IEEE}.
\newblock


\bibitem[\protect\citeauthoryear{Rosenbaum}{Rosenbaum}{2002}]%
        {rosenbaum02observational}
\bibfield{author}{\bibinfo{person}{Paul~R Rosenbaum}.}
  \bibinfo{year}{2002}\natexlab{}.
\newblock \showarticletitle{Observational studies}.
\newblock In \bibinfo{booktitle}{{\em Observational Studies}}.
  \bibinfo{publisher}{Springer}, \bibinfo{pages}{1--17}.
\newblock


\bibitem[\protect\citeauthoryear{Rubin}{Rubin}{2001}]%
        {rubin01using}
\bibfield{author}{\bibinfo{person}{Donald~B Rubin}.}
  \bibinfo{year}{2001}\natexlab{}.
\newblock \showarticletitle{Using propensity scores to help design
  observational studies: application to the tobacco litigation}.
\newblock \bibinfo{journal}{{\em Health Services and Outcomes Research
  Methodology\/}} \bibinfo{volume}{2}, \bibinfo{number}{3}
  (\bibinfo{year}{2001}), \bibinfo{pages}{169--188}.
\newblock


\bibitem[\protect\citeauthoryear{Russakovsky, Deng, Su, Krause, Satheesh, Ma,
  Huang, Karpathy, Khosla, Bernstein, Berg, and Fei-Fei}{Russakovsky
  et~al\mbox{.}}{2015}]%
        {ilscrvc}
\bibfield{author}{\bibinfo{person}{Olga Russakovsky}, \bibinfo{person}{Jia
  Deng}, \bibinfo{person}{Hao Su}, \bibinfo{person}{Jonathan Krause},
  \bibinfo{person}{Sanjeev Satheesh}, \bibinfo{person}{Sean Ma},
  \bibinfo{person}{Zhiheng Huang}, \bibinfo{person}{Andrej Karpathy},
  \bibinfo{person}{Aditya Khosla}, \bibinfo{person}{Michael Bernstein},
  \bibinfo{person}{Alexander~C. Berg}, {and} \bibinfo{person}{Li Fei-Fei}.}
  \bibinfo{year}{2015}\natexlab{}.
\newblock \showarticletitle{{ImageNet Large Scale Visual Recognition
  Challenge}}.
\newblock \bibinfo{journal}{{\em International Journal of Computer Vision\/}}
  \bibinfo{volume}{115}, \bibinfo{number}{3} (\bibinfo{year}{2015}).
\newblock
\showDOI{%
\url{https://doi.org/10.1007/s11263-015-0816-y}}


\bibitem[\protect\citeauthoryear{Russell}{Russell}{1994}]%
        {russell94universal}
\bibfield{author}{\bibinfo{person}{James~A. Russell}.}
  \bibinfo{year}{1994}\natexlab{}.
\newblock \showarticletitle{Is there universal recognition of emotion from
  facial expression? A review of the cross-cultural studies}.
\newblock \bibinfo{journal}{{\em Psychological Bulletin\/}}
  \bibinfo{volume}{115} (\bibinfo{year}{1994}), \bibinfo{pages}{102--141}.
\newblock


\bibitem[\protect\citeauthoryear{Schifanella, Barrat, Cattuto, Markines, and
  Menczer}{Schifanella et~al\mbox{.}}{2010}]%
        {schifanella10folks}
\bibfield{author}{\bibinfo{person}{Rossano Schifanella}, \bibinfo{person}{Alain
  Barrat}, \bibinfo{person}{Ciro Cattuto}, \bibinfo{person}{Benjamin Markines},
  {and} \bibinfo{person}{Filippo Menczer}.} \bibinfo{year}{2010}\natexlab{}.
\newblock \showarticletitle{Folks in Folksonomies: Social Link Prediction from
  Shared Metadata}. In \bibinfo{booktitle}{{\em WSDM}}.
  \bibinfo{publisher}{ACM}.
\newblock
\showISBNx{978-1-60558-889-6}
\showDOI{%
\url{https://doi.org/10.1145/1718487.1718521}}


\bibitem[\protect\citeauthoryear{Schifanella, Redi, and Aiello}{Schifanella
  et~al\mbox{.}}{2015}]%
        {schifanella2015image}
\bibfield{author}{\bibinfo{person}{Rossano Schifanella},
  \bibinfo{person}{Miriam Redi}, {and} \bibinfo{person}{Luca~Maria Aiello}.}
  \bibinfo{year}{2015}\natexlab{}.
\newblock \showarticletitle{An Image Is Worth More than a Thousand Favorites:
  Surfacing the Hidden Beauty of Flickr Pictures}. In \bibinfo{booktitle}{{\em
  ICWSM}}. \bibinfo{publisher}{AAAI}.
\newblock


\bibitem[\protect\citeauthoryear{Shalizi and Thomas}{Shalizi and
  Thomas}{2011}]%
        {shalizi11homophily}
\bibfield{author}{\bibinfo{person}{Cosma~Rohilla Shalizi} {and}
  \bibinfo{person}{Andrew~C Thomas}.} \bibinfo{year}{2011}\natexlab{}.
\newblock \showarticletitle{Homophily and contagion are generically confounded
  in observational social network studies}.
\newblock \bibinfo{journal}{{\em Sociological methods and research\/}}
  \bibinfo{volume}{40}, \bibinfo{number}{2} (\bibinfo{year}{2011}).
\newblock


\bibitem[\protect\citeauthoryear{Siahaan, Redi, and Hanjalic}{Siahaan
  et~al\mbox{.}}{2013}]%
        {redi2014beauty}
\bibfield{author}{\bibinfo{person}{Ernestasia Siahaan},
  \bibinfo{person}{Judith~Alice Redi}, {and} \bibinfo{person}{Alan Hanjalic}.}
  \bibinfo{year}{2013}\natexlab{}.
\newblock \showarticletitle{Beauty is in the scale of the beholder: a
  comparison of methodologies for the subjective assessment of image aesthetic
  appeal}. In \bibinfo{booktitle}{{\em Proceedings of the 2nd ACM international
  workshop on Crowdsourcing for multimedia}}. ACM, \bibinfo{pages}{29--34}.
\newblock


\bibitem[\protect\citeauthoryear{Song, Redi, Vallmitjana, and Jaimes}{Song
  et~al\mbox{.}}{2016}]%
        {song16click}
\bibfield{author}{\bibinfo{person}{Yale Song}, \bibinfo{person}{Miriam Redi},
  \bibinfo{person}{Jordi Vallmitjana}, {and} \bibinfo{person}{Alejandro
  Jaimes}.} \bibinfo{year}{2016}\natexlab{}.
\newblock \showarticletitle{To Click or Not To Click: Automatic Selection of
  Beautiful Thumbnails from Videos}. In \bibinfo{booktitle}{{\em CIKM}}.
  \bibinfo{publisher}{ACM}, \bibinfo{address}{New York, NY, USA},
  \bibinfo{pages}{659--668}.
\newblock
\showISBNx{978-1-4503-4073-1}
\showDOI{%
\url{https://doi.org/10.1145/2983323.2983349}}


\bibitem[\protect\citeauthoryear{Stuart}{Stuart}{2010}]%
        {stuart10matching}
\bibfield{author}{\bibinfo{person}{Elizabeth~A. Stuart}.}
  \bibinfo{year}{2010}\natexlab{}.
\newblock \showarticletitle{Matching Methods for Causal Inference: A Review and
  a Look Forward}.
\newblock \bibinfo{journal}{{\it Statist. Sci.}} \bibinfo{volume}{25},
  \bibinfo{number}{1} (\bibinfo{date}{02} \bibinfo{year}{2010}).
\newblock


\bibitem[\protect\citeauthoryear{Su, Sharma, and Goel}{Su
  et~al\mbox{.}}{2016}]%
        {su16effect}
\bibfield{author}{\bibinfo{person}{Jessica Su}, \bibinfo{person}{Aneesh
  Sharma}, {and} \bibinfo{person}{Sharad Goel}.}
  \bibinfo{year}{2016}\natexlab{}.
\newblock \showarticletitle{The Effect of Recommendations on Network
  Structure}. In \bibinfo{booktitle}{{\em WWW}}.
  \bibinfo{publisher}{International World Wide Web Conferences Steering
  Committee}, \bibinfo{address}{Republic and Canton of Geneva, Switzerland},
  11.
\newblock
\showISBNx{978-1-4503-4143-1}
\showDOI{%
\url{https://doi.org/10.1145/2872427.2883040}}


\bibitem[\protect\citeauthoryear{Susarla, Oh, and Tan}{Susarla
  et~al\mbox{.}}{2012}]%
        {susarla12social}
\bibfield{author}{\bibinfo{person}{Anjana Susarla}, \bibinfo{person}{Jeong-Ha
  Oh}, {and} \bibinfo{person}{Yong Tan}.} \bibinfo{year}{2012}\natexlab{}.
\newblock \showarticletitle{Social networks and the diffusion of user-generated
  content: Evidence from YouTube}.
\newblock \bibinfo{journal}{{\em Information Systems Research\/}}
  \bibinfo{volume}{23}, \bibinfo{number}{1} (\bibinfo{year}{2012}).
\newblock


\bibitem[\protect\citeauthoryear{Thomee, Shamma, Friedland, Elizalde, Ni,
  Poland, Borth, and Li}{Thomee et~al\mbox{.}}{2016}]%
        {thomee2016yfcc100m}
\bibfield{author}{\bibinfo{person}{Bart Thomee}, \bibinfo{person}{David~A
  Shamma}, \bibinfo{person}{Gerald Friedland}, \bibinfo{person}{Benjamin
  Elizalde}, \bibinfo{person}{Karl Ni}, \bibinfo{person}{Douglas Poland},
  \bibinfo{person}{Damian Borth}, {and} \bibinfo{person}{Li-Jia Li}.}
  \bibinfo{year}{2016}\natexlab{}.
\newblock \showarticletitle{{YFCC100M}: The new data in multimedia research}.
\newblock \bibinfo{journal}{{\it Commun. ACM}} \bibinfo{volume}{59},
  \bibinfo{number}{2} (\bibinfo{year}{2016}).
\newblock


\bibitem[\protect\citeauthoryear{Tibshirani, Walther, and Hastie}{Tibshirani
  et~al\mbox{.}}{2001}]%
        {tibshirani2001estimating}
\bibfield{author}{\bibinfo{person}{Robert Tibshirani},
  \bibinfo{person}{Guenther Walther}, {and} \bibinfo{person}{Trevor Hastie}.}
  \bibinfo{year}{2001}\natexlab{}.
\newblock \showarticletitle{Estimating the number of clusters in a data set via
  the gap statistic}.
\newblock \bibinfo{journal}{{\em Journal of the Royal Statistical Society\/}}
  \bibinfo{volume}{63}, \bibinfo{number}{2} (\bibinfo{year}{2001}).
\newblock
\showISSN{1467-9868}
\showDOI{%
\url{https://doi.org/10.1111/1467-9868.00293}}


\bibitem[\protect\citeauthoryear{Wu, Hu, and Gao}{Wu et~al\mbox{.}}{2011}]%
        {wu2011learning}
\bibfield{author}{\bibinfo{person}{Ou Wu}, \bibinfo{person}{Weiming Hu}, {and}
  \bibinfo{person}{Jun Gao}.} \bibinfo{year}{2011}\natexlab{}.
\newblock \showarticletitle{Learning to predict the perceived visual quality of
  photos}. In \bibinfo{booktitle}{{\em ICCV}}. IEEE.
\newblock


\bibitem[\protect\citeauthoryear{Yanai and Qiu}{Yanai and Qiu}{2009}]%
        {yanai09mining}
\bibfield{author}{\bibinfo{person}{Keiji Yanai} {and} \bibinfo{person}{Bingyu
  Qiu}.} \bibinfo{year}{2009}\natexlab{}.
\newblock \showarticletitle{Mining Cultural Differences from a Large Number of
  Geotagged Photos}. In \bibinfo{booktitle}{{\em Proceedings of the 18th
  International Conference on World Wide Web}} {\em (\bibinfo{series}{WWW
  '09})}. \bibinfo{publisher}{ACM}, \bibinfo{address}{New York, NY, USA},
  \bibinfo{pages}{1173--1174}.
\newblock
\showISBNx{978-1-60558-487-4}
\showDOI{%
\url{https://doi.org/10.1145/1526709.1526914}}


\bibitem[\protect\citeauthoryear{Ye, Li, Newman, Adams, and Wang}{Ye
  et~al\mbox{.}}{2017}]%
        {ye2017probabilistic}
\bibfield{author}{\bibinfo{person}{Jianbo Ye}, \bibinfo{person}{Jia Li},
  \bibinfo{person}{Michelle~G Newman}, \bibinfo{person}{Reginald~B Adams},
  {and} \bibinfo{person}{James~Z Wang}.} \bibinfo{year}{2017}\natexlab{}.
\newblock \showarticletitle{Probabilistic Multigraph Modeling for Improving the
  Quality of Crowdsourced Affective Data}.
\newblock \bibinfo{journal}{{\em IEEE Transactions on Affective Computing\/}}
  (\bibinfo{year}{2017}).
\newblock


\bibitem[\protect\citeauthoryear{Zhou, Redi, Haines, and Lalmas}{Zhou
  et~al\mbox{.}}{2016}]%
        {zhou16predicting}
\bibfield{author}{\bibinfo{person}{Ke Zhou}, \bibinfo{person}{Miriam Redi},
  \bibinfo{person}{Andrew Haines}, {and} \bibinfo{person}{Mounia Lalmas}.}
  \bibinfo{year}{2016}\natexlab{}.
\newblock \showarticletitle{Predicting Pre-click Quality for Native
  Advertisements}. In \bibinfo{booktitle}{{\em WWW}}.
  \bibinfo{publisher}{International World Wide Web Conferences Steering
  Committee}, \bibinfo{address}{Republic and Canton of Geneva, Switzerland},
  \bibinfo{pages}{299--310}.
\newblock
\showISBNx{978-1-4503-4143-1}
\showDOI{%
\url{https://doi.org/10.1145/2872427.2883053}}


\end{thebibliography}
\end{document}